\documentclass[reprint,amsmath,amssymb,aps,longbibliography,superscriptaddress,prl]{revtex4-1}

\usepackage[colorlinks=true,linkcolor=blue,citecolor=blue,urlcolor=blue]{hyperref}
\usepackage{graphicx}
\usepackage{dcolumn}
\usepackage{txfonts}
\usepackage{bm}
\usepackage{physics}
\usepackage{siunitx}
\usepackage[T1]{fontenc}
\usepackage[utf8]{inputenc}
\begin{document}

\title{Soliton interferometry with very narrow barriers obtained from spatially dependent dressed states}

\author{Callum L. Grimshaw}
\affiliation{Joint Quantum Centre (JQC) Durham--Newcastle, Department of Physics, Durham University, Durham DH1 3LE, UK}
\author{Thomas P. Billam}
\affiliation{Joint Quantum Centre (JQC) Durham--Newcastle, School of Mathematics, Statistics and Physics, 
\\
Newcastle University, Newcastle upon Tyne NE1 7RU, UK}% Quantum 
\author{Simon A. Gardiner} 
\affiliation{Joint Quantum Centre (JQC) Durham--Newcastle, Department of Physics, Durham University, Durham DH1 3LE, UK}

\date{\today}

\begin{abstract}
Bright solitons in atomic Bose--Einstein condensates are strong candidates
for high precision matter-wave interferometry, as their inherent stability
against dispersion supports long interrogation times. %One 
An analog to a beam splitter is then a narrow potential barrier. A very narrow barrier is desirable for interferometric purposes, but in a  typical realisation
using a blue-detuned optical dipole potential, the width is limited by the 
laser wavelength. We investigate a soliton interferometry scheme using the geometric scalar potential experienced by atoms in a spatially dependent dark state to overcome this limit.  We propose a possible implementation and
numerically probe the effects of deviations from the ideal configuration.
\end{abstract}

\maketitle

Bright solitons are well-known within one-dimensional mean-field models
of elongated attractively-interacting Bose--Einstein condensates (BECs). They have been realized \cite{StreckerFormation2002,KhaykovichFormation2002,CornishFormation2006,LepoutreProduction2016,MeznarsicCesium2019,DiCarliExcitation2019} in BECs of several species \cite{solitary_wave},
and have much-discussed potential for atomic interferometry
\cite{interf1,interf2,interf3,interf4,interf5,interf6,interf7,interf8,interf9,HaineQuantum2018},
owing to long interrogation times enabled by their self-support
against dispersion, and to the phase-sensitivity of soliton collisions
\cite{NguyenCollisions2014}.
Colliding solitons with potential barriers is
a convenient method to create two phase-coherent solitons, and to
recombine two solitons into a phase-dependent output, forming the essential
elements of an interferometer. In the limit of high collisional velocity, and
a barrier narrow relative to the soliton width, a single incident
soliton splits into two solitons with well-defined relative phase
\cite{interf3,interf4,interf5}. Under
the same conditions two solitons colliding ``head-on'' at a barrier recombine with output populations dependent on the incident solitons' relative phase \cite{interf3,interf4}.
These splitting and recombination processes have recently been investigated
experimentally \cite{exp_dur};  
in a typical setup, focused blue-detuned laser beams realize barriers on the micron scale, comparable to a typical soliton width 
\cite{MarchantControlled2013, exp_dur}. How to generate narrower potential barriers required for optimal interferometry remains an important question. A known method to produce subwavelength features is via rapid change over a small region of the amplitude of one of two near-resonant laser fields in an atomic $\Lambda$ configuration, which can be understood in terms of effective potentials experienced by spatially dependent dressed states \cite{ds_tech_1,ds_tech_2,ds_tech_3,ds_tech_4,ds_tech_5,ds_tech_6,ds_tech_7,ds_tech_8,ds_tech_9}.
We propose a technique exploiting these properties to create
a single narrow barrier for soliton interferometry within a quasi-one-dimensional (quasi-1D) BEC. We subject
our proposal to detailed numerical analysis
of both the full $\Lambda$-system and an
effective single-state model, showing it to
provide potentially excellent interferometric performance within an experimentally reasonable regime.

\begin{figure}
    \centering
    \includegraphics[width=\linewidth,trim=0.05cm 0.05cm 0.05cm 0.05cm,clip=true]{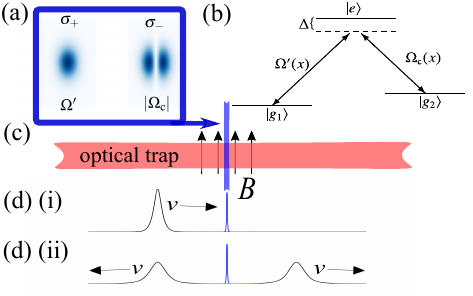}
    \caption{
    Proposed coherent soliton splitting scheme.
    (a) Profiles of differently polarized $\Omega'$ and $\Omega_{\mathrm{c}}$ barrier-forming beams in the $x,z$ plane at $y=0$. (b) Atomic level configuration; we typically consider $\Delta=0$. 
    (c) Schematic of an optical waveguide used to contain the solitons; a magnetic bias field $\bm{B}$ parallel to the $\Omega'$ and $\Omega_{\mathrm{c}}$ beams provides a quantization axis. (d)(i) An initial velocity $v$ soliton propagating in the $+x$ direction
    (d)(ii) splits into two equal-size counterpropagating solitons.
    }
    \label{scheme_diagram}
\end{figure}

We require three internal (hyperfine) 
atomic states, labelled $\ket{g_1}$, $\ket{g_2}$, and $\ket{e}$ in order of increasing energy,
coupled in a $\Lambda$ configuration.
We consider on-resonant  couplings
and neglect spontaneous decay from $\ket{e}$. 
The appropriate quasi-1D vector Gross--Pitaevskii equation (GPE) for a BEC of $N$ mass $m$ atoms, 
transversely confined by a tight harmonic trapping potential of angular frequency $\omega_{r}$, is
\begin{equation}
i \hbar \frac{\partial\psi_j}{\partial t}=-\frac{\hbar ^2}{2m}\frac{\partial ^2\psi_j}{\partial x^2}+\sum_{k}\left(g^{1\mathrm{D}}_{jk}|\psi_{k}|^2\psi_j+\frac{\hbar\Omega_{jk}}{2}\psi_k\right),
\label{VGPE_dim}
\end{equation}
where $j,k\in \{g_1,g_2,e\}$, $g^{1\mathrm{D}}_{jk} = 2\hbar\omega_r
a_{jk}$, the probe beam Rabi frequency
$\Omega_{1e}=\Omega_{e1}=\Omega'(x)$, the control beam Rabi
frequency $\Omega_{2e}=\Omega_{e2}=\Omega_{\mathrm{c}}(x)$, and all other
$\Omega_{jk}=0$. The spatially-dependent coupling leads to a spatially-dependent dressed-state basis in which an artificial gauge
field term appears \cite{ds_tech_1,ds_tech_2,ds_tech_8} in the form of a vector potential $A=iU^{\dagger}\partial_xU$. This results in the geometric scalar potential
\begin{equation}
V(x)= \bra{d} \frac{A^2}{2} \ket{d} = \frac{\hbar^2}{2m}\left(\frac{\Omega' \partial_x \Omega_{\mathrm{c}}-\Omega_{\mathrm{c}} \partial_x \Omega'}{\Omega'^2+\Omega_{\mathrm{c}}^2}\right)^2
\label{potential}
\end{equation}
for the dark state $\ket{d}$.
We illustrate our scheme, using equal-width zeroth- and first-order Hermite--Gaussian modes for the probe and control beams, respectively, in Fig.~\ref{scheme_diagram}.  We express the Rabi frequencies as $\Omega'(x) = \Omega_0 l^{1/2} \mathop{\phi_0(x)}$ and
$\Omega_{\mathrm{c}}(x) = \Omega_1 l^{1/2} \mathop{\phi_1(x)}$, where $\phi_0(x)=[
2/(\pi^{1/2}l)]^{1/2}\mathop{\mathrm{exp}(-x^2/l^2)}$ and $\phi_1(x)=
(2x/l)\mathop{\phi_0(x)}$ are normalized
Hermite--Gaussian functions of
width $l$. Crucially, $\Omega_{\mathrm{c}}(x)= \mathop{h(x)}\mathop{\Omega'(x)}$, where $h(x) = x/w$ and $w= (l/2)(\Omega_0/\Omega_1)$. In physical terms,
$w=(\delta l/2)(P_0/P_1)^{1/2}$, where $\delta$ is the 
($\simeq 1$)
ratio
between dipole transition matrix elements, and $P_n$
the $n^{\mathrm{th}}$-order Hermite--Gaussian beam power. The common envelope function then cancels in the resulting dark state $\ket{d} =[\ket{g_{1}} - (w/x)\ket{g_{2}}]/[1+(w/x)^{2}]^{1/2}$ and [via Eq.~(\ref{potential})] the geometric scalar potential
\begin{equation}
V_h(x)=\frac{\hbar^{2}}{2mw^2}\frac{1}{[1+(x/w)^2]^2}.
\label{V_h}
\end{equation}
Phase-locking of the two laser beams is critical (to avoid population of bright states) when $|g_{2}\rangle$ contributes significantly to $|d\rangle$ [see Fig.~\ref{fig_dyna}(c)], however techniques for phase-stable Raman coupling of hyperfine states are well established \cite{Zhao2020,Arias2017,Rosi2014}, The $\Omega_{\mathrm{c}}$ beam can be generated using an essentially noise-free passive phase retarder \cite{HG_trap}, or DMD \cite{Zupancic2016}, and changes in
optical path length between the two beams (potentially leading to phase drift) can be interferometrically stabilized if required \cite{Uehlinger2013}. Active stabilization techniques \cite{as_trap} can be used in co-locating the beams,
noting that slightly unequal beam centres and widths (relative to $l$) do not cause significant qualitative change within the
relevant regime of decreasing $w$. 

Far from the barrier the dark state approaches $\ket{g_{1}}$, and we initialize with a soliton in this internal state.  Slow (relative to
internal state dynamics) passage across the barrier minimizes coupling to other dressed states; the dark state $\ket{d}$ is adiabatically followed, and the excited state $\ket{e}$ remains unpopulated, preventing spontaneous decay.  
This is compatible with
the ``sudden'' passage required for interferometrically desirable high-velocity and narrow-barrier collisions, 
as we can choose $\Omega\equiv \Omega'(0)
= (2/\pi^{1/2})^{1/2} \Omega_{0}
$, setting the timescale for internal atomic dynamics \textit{independently\/} from the value of $w$.
It is in principle always possible to set $\Omega$ sufficiently high to ensure internal dynamics faster than 
passage across the barrier.
An approximate single-state model,
assuming the atoms remain
in the internal dark state with spatial profile $\psi_d$, leads 
to the scalar GPE
\begin{equation}
i \hbar \frac{\partial \psi_d}{\partial t} = \left( - \frac{\hbar^2}{2m} \frac{\partial^2}{\partial x^2} + V_d + g_{11}^\mathrm{1D} | \psi_d|^2 \right) \psi_d. \label{GPE_dim}
\end{equation}
In the idealized scenario that the scattering lengths $a_{jk}$ are all equal,
Eq.~(\ref{GPE_dim}) applies with $V_d = V_h$, consistent with
in this case bright soliton solutions to
Eq.~(\ref{VGPE_dim}) existing 
with spatial density profile independent of the internal state population distribution \cite{int_vib,manakov1}. A more realistic scenario
is to tune $a_{11}$ by a
Feshbach resonance to a negative value to create bright solitons in state
$\ket{g_1}$, where we assume the other scattering lengths are fixed at a background
value $a_{jk} = g a_{11}$, in which case
\begin{equation}
V_d(x,|\psi_d|^{2}) = 
V_h(x) +  (g-1)\frac{2(x/w)^2+1}{[1+(x/w)^2]^2}
g_{11}^\mathrm{1D}|\psi_d|^2,
\label{eq_nl_barr}
\end{equation}
reverting to $V_d= V_h$ when $g=1$.

\begin{figure}[t!]
\includegraphics[width=\linewidth,trim=0.8cm 0.0cm 4.0cm 0.5cm,clip=true]{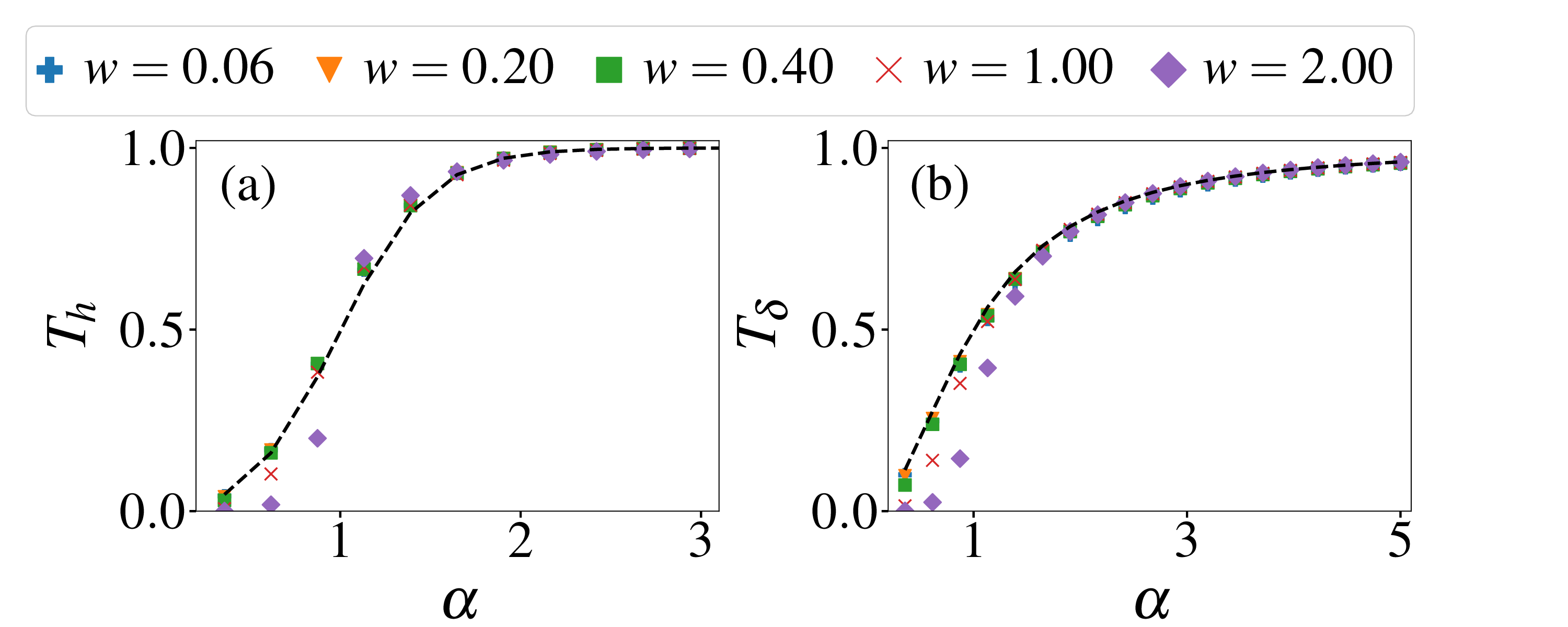}
\caption{Bright soliton collisions with the geometric barrier $V_h$ [Eq.~(\ref{potential})] in the
scalar GPE [Eq.~(\ref{GPE_dim})]. The plots show %the 
transmission as a function of $\alpha$ %(the
(ratio between velocity and 
barrier area, in units of $\hbar^{-1}$), for the barriers $V_h$ (a) and
$V_{\delta}$ (b), and different values of the width $w$. Dashed lines in (a)
and (b) show 
%the 
high-velocity limits for barriers $V_{\mathrm{RM}}$ and
$V_{\delta}$, respectively.}
\label{lorentzian_delta_comparison}
\end{figure}

We simulate the
vector GPE [Eq.~(\ref{VGPE_dim})] and scalar GPE [Eq.~(\ref{GPE_dim})] with
periodic boundary conditions, corresponding to a quasi-1D ring trap
configuration.  We take  $^{85}$Rb with $\ket{g_1} =
\ket{F=2,M_F-2}$ and $\ket{g_2} = \ket{F=2,M_F=0}$ coupled via the D1 line as
an inspirational example.  This 
has a wide Feshbach resonance around
$B_0=156$~G~\cite{feshbachresonance_1,feshbachresonance_2}, which we
use to tune $a_{11} \approx \SI{-12}{\bohr}$, within
the stable soliton region \cite{CornishFormation2006,stab_2,stab_3}.  Assuming
all other scattering lengths 
to be equal to the background value
$a_{\mathrm{bg}}=\SI{-441}{\bohr}$ yields $g\approx 40$. To broaden our
analysis, we vary $g$ between $-40$ and $40$.  We work in ``soliton'' units of
length $\hbar ^{2}/m|g^{1\mathrm{D}}_{11}|N$, time $\hbar
^{3}/m(g^{1\mathrm{D}}_{11}N)^{2}$, and energy $m(g^{1\mathrm{D}}_{11}N/\hbar
)^{2}$ \cite{dimensionless_variables}. 
Unless otherwise stated, we express
quantities in these units, with total density normalized to 1.  
We set $l = 2\sqrt{2}$ 
in our vector GPE simulations; for the above value of $a_{11}$, %with 
$N=2500$ and $\omega_r=2\pi\times\SI{40}{\hertz}$,
this corresponds to an SI value of $l=\SI{2.7}{\um}$ \cite{exp_dur}. 
We assume an initial
bright soliton  $\psi_{1}=
(1/2)\mathop{\mathrm{sech}([x+L/4]/2)}e^{ivx}$ in state $\ket{g_1}$, with $\psi_{2}= \psi_{e}=
0$, and ring trap circumference $L= 64\pi$. 

\begin{figure}[t!]
\includegraphics[width=\linewidth,trim=1.3cm 0.5cm 4cm 1cm,clip=true]{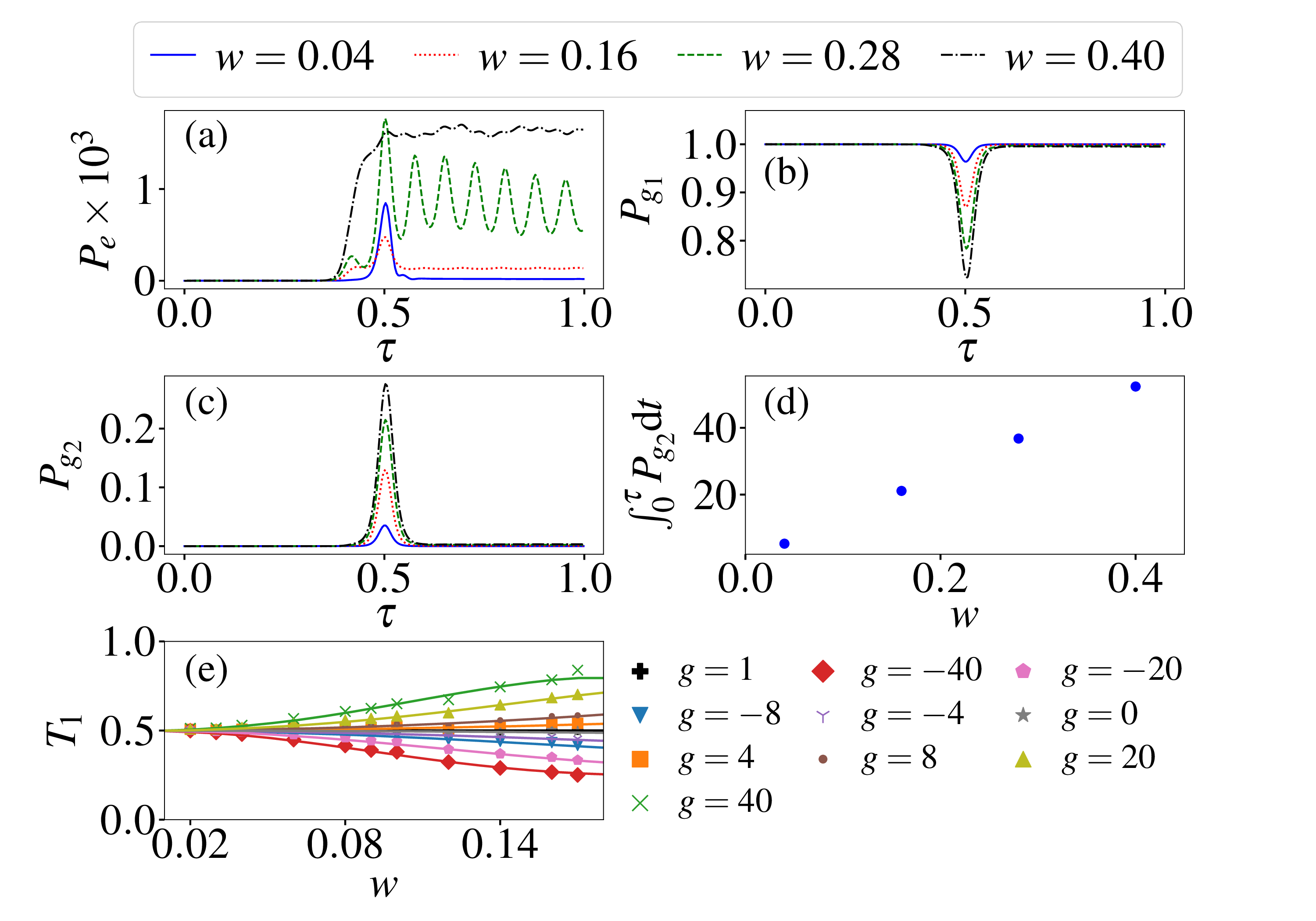}
\caption{Bright soliton collisions with the proposed barrier configuration in the vector GPE [Eq.~(\ref{VGPE_dim})].
(a--c) Populations as functions of time 
(in units of $\tau=L/2v$, the time over which
the soliton moves from $-L/4$ to $L/4$) of states $\ket{e}$, $\ket{g_1}$, and $\ket{g_2}$, respectively. (d) Integrated time spent in
state $\ket{g_2}$ as a function of $w$.  In (a--d), we set
$\Omega= 10^4$ and $g= 1$. (e) Transmission as a function of $w$ for different
values of $g$, where we set $\Omega= 10^6$, and fix incoming soliton
velocities at values resulting in $T=0.5$ for scalar GPE
simulations with $V_h$ barriers [Fig.~\ref{lorentzian_delta_comparison}
(a)]; solid lines show equivalent-parameter scalar GPE simulations with fully nonlinear barrier
$V_d(x,|\psi_{d}|^{2})$ [Eq.~(\ref{eq_nl_barr})].}
\label{fig_dyna}
\end{figure}

We first use the scalar GPE [Eq.~(\ref{GPE_dim})] 
to investigate soliton collisions with the
squared-Lorentzian barrier $V_h$. We compare the total fraction of transmitted atoms $T$
with the analytic approximation for collisions with a 
same-height-and-area
Rosen--Morse barrier, 
$V_{\mathrm{RM}}= [1/(2w^2)]\mathop{\mathrm{sech}^2(4x/[\pi w])} $,
in the high-velocity limit
(neglecting the nonlinear term 
during the collision) \cite{exp_dur, landauandlifshitz}. We also compare indirectly to scalar GPE simulations with a same-area
$\delta$-function barrier, 
$V_d = V_{\delta}(x)= [\pi/(4w)]\mathop{\delta(x)}$, which approach their
own analytic high-velocity limit $T_{\delta}(\alpha)=
\alpha^2/(1+\alpha^2)$, where $\alpha=4vw/\pi$ is the rato between velocity and
barrier area~\cite{dfanalytics1}. Figure \ref{lorentzian_delta_comparison}
shows numerical
transmission curves for $V_h$ and $V_{\delta}$ barriers with different values of $w$ plotted against 
$\alpha$. As $w$ decreases, the transmission
curves approach the analytic high-velocity limits for  $V_{\mathrm{RM}}$ and
$V_{\delta}$.
How Fig.~\ref{lorentzian_delta_comparison}(a) and (b) differ
illustrates an important point. 
Within an interferometer, an effective soliton beamsplitter should 
achieve $T=0.5$ in the tunneling
regime, where the ratio $\gamma$ between  per-atom kinetic energy and barrier
height satisfies $\gamma < 1$; the outgoing
soliton velocities may otherwise have significantly different magnitudes owing to velocity filtering~\cite{exp_dur}. As collision velocities increase, 
we need decreasing barrier widths
to remain in the tunneling regime
\cite{interf5,tunneling2,si1}. The $V_h$ barrier width $w$ and area
$\pi/(4w)$ are intrinsically inversely related, \textit{fixing\/} the ratio
$\gamma= (vw)^2 = \left( \pi/4 \right)^2\alpha^2$. Assuming $T=0.5$  
occurs close to $\alpha=1$, 
$\gamma$ tends towards $(\pi/4)^2
\approx 0.61$. 
The $\delta$-function limit $\gamma
\rightarrow 0$ 
is therefore not attained
with the $V_h$ barrier;
as the width decreases with increasing ratio
$\Omega_1/\Omega_0$, the velocity at which $T=0.5$ 
is nonetheless within the $\gamma < 1$ tunneling regime.
In Fig.~\ref{fig_dyna} we investigate these same collisions using the vector
GPE description [Eq.~(\ref{VGPE_dim})] for varying $w$. We fix incoming
soliton velocities at values resulting in $T=0.5$ for 
the scalar GPE with $V_d = V_h$ [Fig.~\ref{lorentzian_delta_comparison}(a)].  In Fig.~\ref{fig_dyna}(a--d) we consider equal scattering lengths
($g=1$) and characterize 
internal state populations 
as functions of time during the collision, showing
the integrated time spent in state $\ket{g_2}$ 
as a function of $w$ in (d). As expected, decreasing 
$w$ generally reduces the populations of $\ket{g_2}$ and $\ket{e}$
and increases that of $\ket{g_1}$; the integrated time spent
in state $\ket{g_2}$ also decreases. In Fig.~\ref{fig_dyna}(e) we show the
transmission $T$ as a function of $w$ for a range of scattering length ratios
$g$; as $w$ decreases, the effects of $g \ne 1$ reduce.
The solid lines in Fig.~\ref{fig_dyna}(e) show results 
of the scalar GPE with fully nonlinear $V_d(x,|\psi_d|^{2})$ [Eq.~(\ref{eq_nl_barr})],
which clearly matches the vector GPE well over the range of $g$ we explore.

\begin{figure}[t!]
\includegraphics[width=\linewidth,trim=0.8cm 0cm 4cm 0.75cm,clip=true]{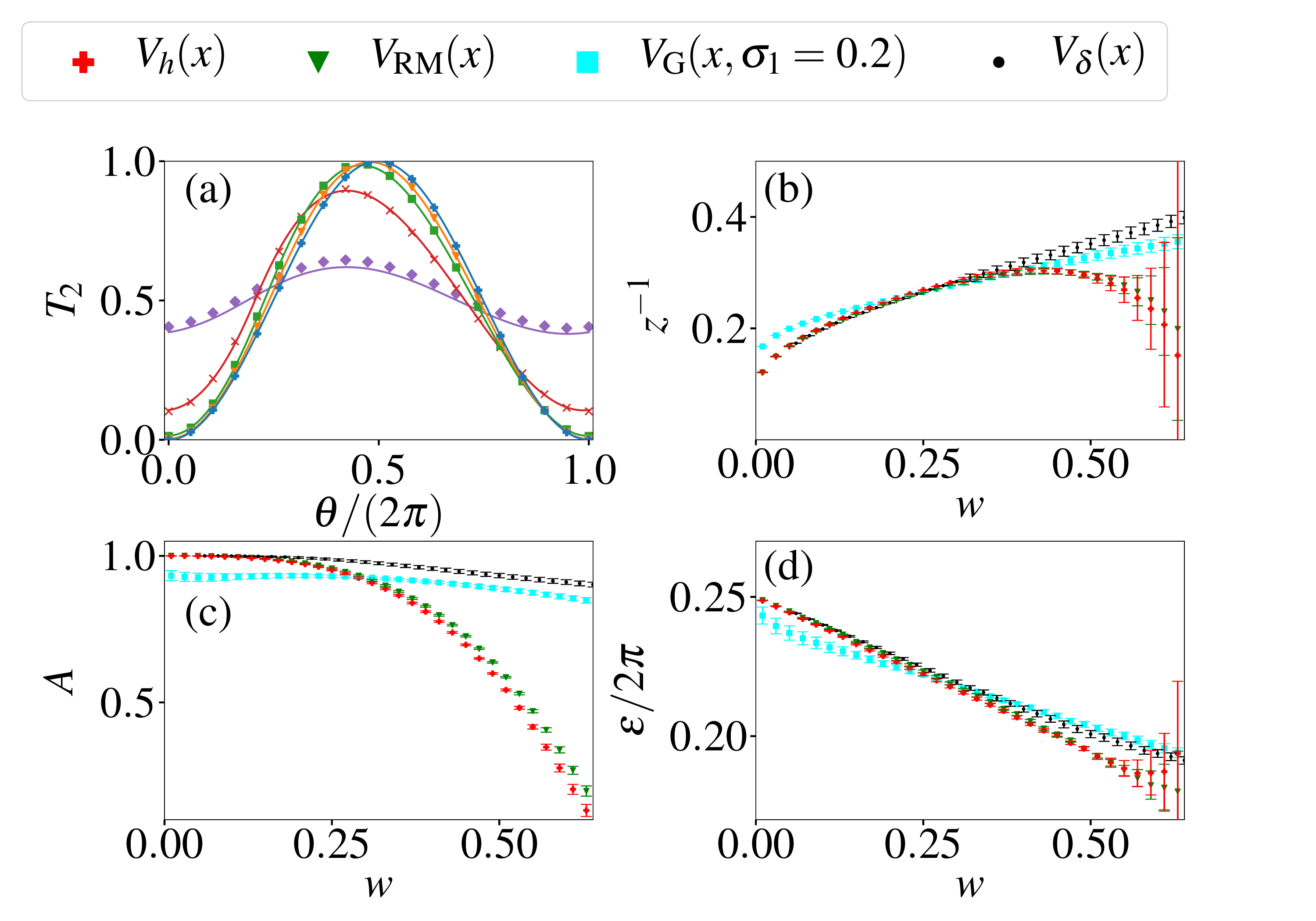}
\caption{
Bright soliton interferometry with the geometric barrier $V_h$ in the scalar
GPE. (a) Transmission at recombination $T_2$ against imposed 
phase
$\theta$ for $w= 0.01$ (blue plus), $w= 0.1$ (orange triangle), $w=
0.2$ (green cross), $w= 0.4$ (red square), and $w= 0.6$ (purple
diamond). (b)--(d) Fitted values [using Eq.~(\ref{eq_int_skew})] of $z^{-1}$, $A$, and $\varepsilon$, respectively, for $V_h$ and for the alternative barrier shapes $V_{\mathrm{RM}}$, $V_{\delta}$, and
$V_{\mathrm{G}}(\sigma=0.2)$ (see text) for varying $w$.}
\label{fig_int_params}
\end{figure}

While various interferometric configurations are possible, we consider a
conceptually simple quasi-1D ring trap with a single barrier. The barrier splits a single soliton into two equal-amplitude, equal-speed
counterpropagating daughter solitons, which pass through one another and subsequently phase-sensitively recombine at the same barrier \cite{interf7}. Imposing a relative phase $\theta$ between the daughter
solitons,
the fraction of atoms recombined to one side of the barrier 
%to
should
vary sinusoidally with $\theta$ in the high velocity and narrow barrier limit (i.e., $w \rightarrow
0$). We otherwise expect a nonlinearity-induced ``skew'' in the
sinusoidal dependence 
\cite{interf4}, and employ
(generalized) Clausen functions $S_z(\theta)$
to empirically parametrize 
this effect.
We fit the final
population on the ``transmitted'' side of the barrier after the second
collision (the recombination) with
\begin{equation}
    T_2(\theta)=\frac{1}{2}\left[1+A\mathop{S_{z}(\theta-\varepsilon)}\right],
\label{eq_int_skew}
\end{equation}
which ranges from a sawtooth function  ($z=1$) to a sinusoid ($z\rightarrow \infty$).
To improve fitting convergence and ensure
bounded limits, we fit  and present results in terms of $z^{-1}$, where smaller $z^{-1}$ corresponds to less skew \cite{fitting, SciPy2020}.
The phase shift $\varepsilon$ incorporates relative phases accumulated during barrier collisions and subsequent evolution, and $A$ is the contrast or ``fringe visibility.'' For a $\delta$-function barrier in the high-velocity limit
$z^{-1} \rightarrow 0$, $A= 1$ and $\varepsilon= \pi/2$
\cite{interf4,dfanalytics1}. In Fig.~\ref{fig_int_params} we show
this limit is effectively reached
with the $V_h$ barrier in the scalar GPE. We compare this scenario to
the scalar GPE with alternative barriers $V_d =V_\delta (x)$, 
$V_d = V_\mathrm{RM}(x)$, 
and a narrow, fixed-width, Gaussian barrier with equal area to
$V_h$: $V_d =
V_{\mathrm{G}}(x;\sigma)=\{
[\pi/(4w)]/[(2\pi)^{1/2}\sigma]\}\,\mathrm{exp}
(-x^2/[2\sigma^2])$.  For each data point %we use 
a root-finding
algorithm
sets
the initial velocity to
achieve transmission $T=0.5$ at the first collision, and we model
a range of imposed phases $\theta$.
Figure~\ref{fig_int_params}(a) shows $T_2(\theta)$ for the $V_h$ barrier,
directly illustrating the decrease in skew for decreasing $w$.
Figures~\ref{fig_int_params}(b--d) show the values of $z^{-1}$, $A$, and
$\varepsilon$ extracted by fitting Eq.~(\ref{eq_int_skew}) to the numerical
simulations at width $w$. The $V_h$ barrier, and its Rosen--Morse approximant
$V_\mathrm{RM}$, smoothly approach the ideal high-velocity $\delta$-function
result of $A=1$ and $\varepsilon = \pi / 2$ as $w \rightarrow 0$; note
the fixed-width Gaussian barrier performs better at $w\gtrsim 0.3$,
but cannot smoothly reach this result.
The parameter $z^{-1}$ does not drop smoothly to zero, however 
at $z^{-1} < 0.2$ the skew 
is barely resolved and the 3-parameter fit 
of Eq.~(\ref{eq_int_skew}) effectively over-fits in this limit. 

\begin{figure}[t!]
\includegraphics[width=\linewidth,trim=0.1cm 0.0cm 4cm 0.75cm,clip=true]{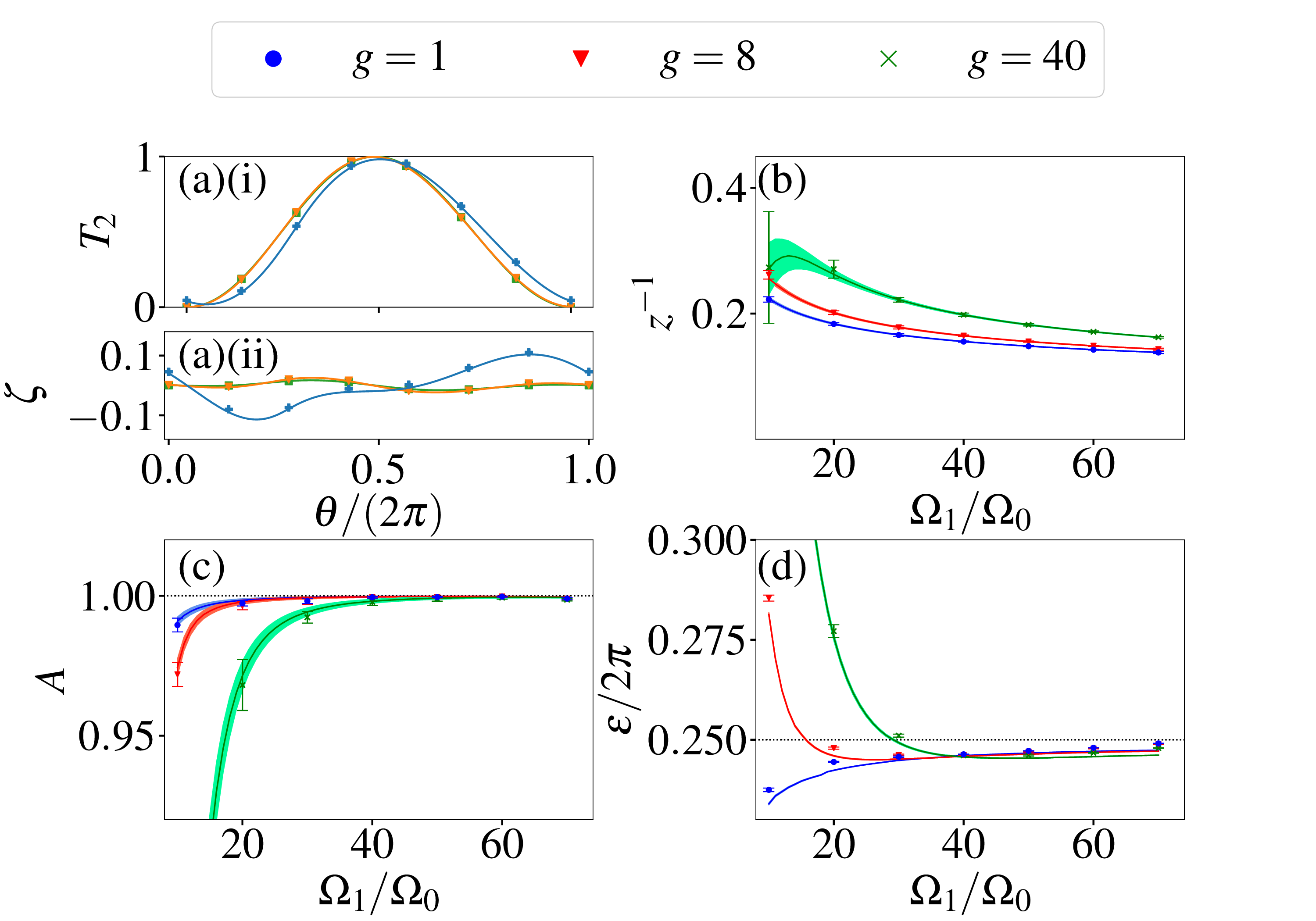}
\caption{
Bright soliton interferometry in the vector GPE [Eq.~(\ref{VGPE_dim})].  (a)(i) Transmission at
recombination $T_2$ against imposed phase $\theta$, 
and (a)(ii)
difference from the ideal sinusoid
$\zeta=T_2-(1/2)[1+\sin(\theta-\pi/2)]$, for
$\Omega_1/\Omega_0= 20$ (blue plus), $\Omega_1/\Omega_0= 40$ (orange
triangle), $\Omega_1/\Omega_0= 60$ (green square), with $g=40$.
(b)--(d)
Values of $z^{-1}$, $A$, and $\varepsilon$, respectively, found by fitting with
Eq.~(\ref{eq_int_skew}) for $g= 1$, $g= 8$, $g=40$. Solid lines show the fit to scalar GPE simulations 
with fully nonlinear
barrier $V_d(x,|\psi|^{2})$ and equivalent parameters (shaded areas indicate error
ranges). We use $\Omega= 10^4$ for $\Omega_1/\Omega_0=$ 10, 20, 30 and
$\Omega= 5\times 10^4$ for $\Omega_1/\Omega_0=$ 40, 50, 60, 70 (see
text).} 
\label{fig_3s}
\end{figure}

In Fig.~\ref{fig_3s} we analyze the interferometer using the vector GPE [Eq.~(\ref{VGPE_dim})] for
$g =1, 8, 40$,
presenting the results as functions of the ratio $\Omega_1 / \Omega_0 \equiv l / (2w)$.
Due to the high computational demands of setting the initial velocities with our previously employed root-finding algorithm, we use values determined for Fig.~\ref{fig_int_params} at equivalent widths $w$ for the $V_h$ barrier. In the ideal limit these velocities are the same,
but 
otherwise 
significantly different $A$ and $\varepsilon$ values result for different $g$. 
Figure~\ref{fig_3s}(a) shows directly the decrease in skew for $g=40$ as
$\Omega_1 / \Omega_0$ increases. Figure~\ref{fig_3s} (b--d) shows how the
fit-extracted parameters $z^{-1}$, $A$, and $\varepsilon$ tend towards
the ideal limit as $\Omega_1 / \Omega_0$ increases. As in Fig.~\ref{fig_dyna},
solid lines show scalar GPE simulations with fully nonlinear potential $V_d(x,|\psi|^{2})$ (shaded areas indicate error ranges from fitting), again
showing excellent agreement.
We require high $\Omega$ values 
to keep internal state dynamics sufficiently fast relative to the collision
duration as $\Omega_1 / \Omega_0$ increases (values used 
are given in the figure caption).  Extension to even higher
$\Omega_1 / \Omega_0$ values is in principle enabled by
raising $\Omega$ further;
the physically desirable strong separation of timescales makes this an increasingly challenging regime to fully simulate, however. Briefly considering off-resonant excitation and spontaneous decay, we similarly note that, in the desired regime of operation, the splitting at the barrier is a predominantly linear effect \cite{dfanalytics1,interf4}. Therefore, as an approximate model, we numerically solve the time-independent, three-state linear scattering problem for an incoming plane wave with wavenumber $k>0$ in state $\ket{g_1}$, and purely outgoing plane waves in every other channel. We include loss due to spontaneous decay from $|e\rangle$ by combining an imaginary term with the detuning,
producing $-\Delta - i \Gamma /2$, where $\Gamma$ is the excited state linewidth. With $^{85}$Rb parameters corresponding to the rightmost points in Fig.~\ref{fig_3s}~(b--d), we find the effects of spontaneous decay and realistic detuning (equal to the linewidth) are negligible at the wavenumber $k$ required for equal splitting \footnote{Specifically, the wavenumber $k$ required for equal splitting shifts by $\approx 2\,$\%, and $\approx 2\,$\% of the incoming atom flux is lost at this wavenumber.}.

We have described a technique to create a single very narrow barrier for soliton interferometry using a geometric scalar potential \cite{ds_tech_1, ds_tech_2}, based on two overlapping Hermite--Gaussian mode laser beams. We have used approximate scalar GPE and full vector GPE models to characterize both splitting and
interferometric recombination at this barrier,
demonstrating
how 
to realize
a very narrow effective barrier
using moderately high laser intensity ratios.
Critically, 
the initial equal splitting of a single soliton is then a tunneling rather than a velocity filtering process, and near-unit
interferometric contrast  is in principle achievable.
We have also shown a 
scalar GPE with correctly chosen nonlinear barrier
potential provides an excellent description of the system, provided the
intensity of the weaker beam is sufficiently high. We hope this proposal
will find practical application in the upcoming generation of 
matter-wave bright soliton experiments.

Additional data related to the findings reported in this paper is made available by source \cite{data}.

We thank 
S. L. Cornish, 
A. Guttridge,
I. G. Hughes 
and 
A. Rakonjac 
for useful discussions. C.L.G. is supported by the UK EPSRC. This work made use of the Durham University Hamilton HPC Service. 

\bibliography{darkstatebarrier}

\begin{thebibliography}{51}%
\makeatletter
\providecommand \@ifxundefined [1]{%
 \@ifx{#1\undefined}
}%
\providecommand \@ifnum [1]{%
 \ifnum #1\expandafter \@firstoftwo
 \else \expandafter \@secondoftwo
 \fi
}%
\providecommand \@ifx [1]{%
 \ifx #1\expandafter \@firstoftwo
 \else \expandafter \@secondoftwo
 \fi
}%
\providecommand \natexlab [1]{#1}%
\providecommand \enquote  [1]{``#1''}%
\providecommand \bibnamefont  [1]{#1}%
\providecommand \bibfnamefont [1]{#1}%
\providecommand \citenamefont [1]{#1}%
\providecommand \href@noop [0]{\@secondoftwo}%
\providecommand \href [0]{\begingroup \@sanitize@url \@href}%
\providecommand \@href[1]{\@@startlink{#1}\@@href}%
\providecommand \@@href[1]{\endgroup#1\@@endlink}%
\providecommand \@sanitize@url [0]{\catcode `\\12\catcode `\$12\catcode
  `\&12\catcode `\#12\catcode `\^12\catcode `\_12\catcode `\%12\relax}%
\providecommand \@@startlink[1]{}%
\providecommand \@@endlink[0]{}%
\providecommand \url  [0]{\begingroup\@sanitize@url \@url }%
\providecommand \@url [1]{\endgroup\@href {#1}{\urlprefix }}%
\providecommand \urlprefix  [0]{URL }%
\providecommand \Eprint [0]{\href }%
\providecommand \doibase [0]{http://dx.doi.org/}%
\providecommand \selectlanguage [0]{\@gobble}%
\providecommand \bibinfo  [0]{\@secondoftwo}%
\providecommand \bibfield  [0]{\@secondoftwo}%
\providecommand \translation [1]{[#1]}%
\providecommand \BibitemOpen [0]{}%
\providecommand \bibitemStop [0]{}%
\providecommand \bibitemNoStop [0]{.\EOS\space}%
\providecommand \EOS [0]{\spacefactor3000\relax}%
\providecommand \BibitemShut  [1]{\csname bibitem#1\endcsname}%
\let\auto@bib@innerbib\@empty
%</preamble>
\bibitem [{\citenamefont {Strecker}\ \emph {et~al.}(2002)\citenamefont
  {Strecker}, \citenamefont {Partridge}, \citenamefont {Truscott},\ and\
  \citenamefont {Hulet}}]{StreckerFormation2002}%
  \BibitemOpen
  \bibfield  {author} {\bibinfo {author} {\bibfnamefont {K.~E.}\ \bibnamefont
  {Strecker}}, \bibinfo {author} {\bibfnamefont {G.~B.}\ \bibnamefont
  {Partridge}}, \bibinfo {author} {\bibfnamefont {A.~G.}\ \bibnamefont
  {Truscott}}, \ and\ \bibinfo {author} {\bibfnamefont {R.~G.}\ \bibnamefont
  {Hulet}},\ }\bibfield  {title} {\enquote {\bibinfo {title} {Formation and
  propagation of matter-wave soliton trains},}\ }\href {\doibase
  10.1038/nature747} {\bibfield  {journal} {\bibinfo  {journal} {Nature}\
  }\textbf {\bibinfo {volume} {417}},\ \bibinfo {pages} {150} (\bibinfo {year}
  {2002})}\BibitemShut {NoStop}%
\bibitem [{\citenamefont {Khaykovich}\ \emph {et~al.}(2002)\citenamefont
  {Khaykovich}, \citenamefont {Schreck}, \citenamefont {Ferrari}, \citenamefont
  {Bourdel}, \citenamefont {Cubizolles}, \citenamefont {Carr}, \citenamefont
  {Castin},\ and\ \citenamefont {Salomon}}]{KhaykovichFormation2002}%
  \BibitemOpen
  \bibfield  {author} {\bibinfo {author} {\bibfnamefont {L.}~\bibnamefont
  {Khaykovich}}, \bibinfo {author} {\bibfnamefont {F.}~\bibnamefont {Schreck}},
  \bibinfo {author} {\bibfnamefont {G.}~\bibnamefont {Ferrari}}, \bibinfo
  {author} {\bibfnamefont {T.}~\bibnamefont {Bourdel}}, \bibinfo {author}
  {\bibfnamefont {J.}~\bibnamefont {Cubizolles}}, \bibinfo {author}
  {\bibfnamefont {L.~D.}\ \bibnamefont {Carr}}, \bibinfo {author}
  {\bibfnamefont {Y.}~\bibnamefont {Castin}}, \ and\ \bibinfo {author}
  {\bibfnamefont {C.}~\bibnamefont {Salomon}},\ }\bibfield  {title} {\enquote
  {\bibinfo {title} {Formation of a matter-wave bright soliton},}\ }\href
  {\doibase 10.1126/science.1071021} {\bibfield  {journal} {\bibinfo  {journal}
  {Science}\ }\textbf {\bibinfo {volume} {296}},\ \bibinfo {pages} {1290}
  (\bibinfo {year} {2002})}\BibitemShut {NoStop}%
\bibitem [{\citenamefont {Cornish}\ \emph {et~al.}(2006)\citenamefont
  {Cornish}, \citenamefont {Thompson},\ and\ \citenamefont
  {Wieman}}]{CornishFormation2006}%
  \BibitemOpen
  \bibfield  {author} {\bibinfo {author} {\bibfnamefont {S.~L.}\ \bibnamefont
  {Cornish}}, \bibinfo {author} {\bibfnamefont {S.~T.}\ \bibnamefont
  {Thompson}}, \ and\ \bibinfo {author} {\bibfnamefont {C.~E.}\ \bibnamefont
  {Wieman}},\ }\bibfield  {title} {\enquote {\bibinfo {title} {Formation of
  bright matter-wave solitons during the collapse of attractive
  {B}ose--{E}instein condensates},}\ }\href {\doibase
  10.1103/PhysRevLett.96.170401} {\bibfield  {journal} {\bibinfo  {journal}
  {Phys. Rev. Lett.}\ }\textbf {\bibinfo {volume} {96}},\ \bibinfo {pages}
  {170401} (\bibinfo {year} {2006})}\BibitemShut {NoStop}%
\bibitem [{\citenamefont {Lepoutre}\ \emph {et~al.}(2016)\citenamefont
  {Lepoutre}, \citenamefont {Fouch\'e}, \citenamefont {Boiss\'e}, \citenamefont
  {Berthet}, \citenamefont {Salomon}, \citenamefont {Aspect},\ and\
  \citenamefont {Bourdel}}]{LepoutreProduction2016}%
  \BibitemOpen
  \bibfield  {author} {\bibinfo {author} {\bibfnamefont {S.}~\bibnamefont
  {Lepoutre}}, \bibinfo {author} {\bibfnamefont {L.}~\bibnamefont {Fouch\'e}},
  \bibinfo {author} {\bibfnamefont {A.}~\bibnamefont {Boiss\'e}}, \bibinfo
  {author} {\bibfnamefont {G.}~\bibnamefont {Berthet}}, \bibinfo {author}
  {\bibfnamefont {G.}~\bibnamefont {Salomon}}, \bibinfo {author} {\bibfnamefont
  {A.}~\bibnamefont {Aspect}}, \ and\ \bibinfo {author} {\bibfnamefont
  {T.}~\bibnamefont {Bourdel}},\ }\bibfield  {title} {\enquote {\bibinfo
  {title} {Production of strongly bound $^{39}\mathrm{K}$ bright solitons},}\
  }\href {\doibase 10.1103/PhysRevA.94.053626} {\bibfield  {journal} {\bibinfo
  {journal} {Phys. Rev. A}\ }\textbf {\bibinfo {volume} {94}},\ \bibinfo
  {pages} {053626} (\bibinfo {year} {2016})}\BibitemShut {NoStop}%
\bibitem [{\citenamefont {{Me{\v{z}}nar{\v{s}}i{\v{c}}}}\ \emph
  {et~al.}(2019)\citenamefont {{Me{\v{z}}nar{\v{s}}i{\v{c}}}}, \citenamefont
  {{Arh}}, \citenamefont {{Brence}}, \citenamefont {{Pi{\v{s}}ljar}},
  \citenamefont {{Gosar}}, \citenamefont {{Gosar}}, \citenamefont
  {{{\v{Z}}itko}}, \citenamefont {{Zupani{\v{c}}}},\ and\ \citenamefont
  {{Jegli{\v{c}}}}}]{MeznarsicCesium2019}%
  \BibitemOpen
  \bibfield  {author} {\bibinfo {author} {\bibfnamefont {T.}~\bibnamefont
  {{Me{\v{z}}nar{\v{s}}i{\v{c}}}}}, \bibinfo {author} {\bibfnamefont
  {T.}~\bibnamefont {{Arh}}}, \bibinfo {author} {\bibfnamefont
  {J.}~\bibnamefont {{Brence}}}, \bibinfo {author} {\bibfnamefont
  {J.}~\bibnamefont {{Pi{\v{s}}ljar}}}, \bibinfo {author} {\bibfnamefont
  {K.}~\bibnamefont {{Gosar}}}, \bibinfo {author} {\bibfnamefont
  {{\v{Z}}}~\bibnamefont {{Gosar}}}, \bibinfo {author} {\bibfnamefont
  {R.}~\bibnamefont {{{\v{Z}}itko}}}, \bibinfo {author} {\bibfnamefont
  {E.}~\bibnamefont {{Zupani{\v{c}}}}}, \ and\ \bibinfo {author} {\bibfnamefont
  {P.}~\bibnamefont {{Jegli{\v{c}}}}},\ }\bibfield  {title} {\enquote {\bibinfo
  {title} {Cesium bright matter-wave solitons and soliton trains},}\ }\href
  {\doibase 10.1103/PhysRevA.99.033625} {\bibfield  {journal} {\bibinfo
  {journal} {Phys. Rev. A}\ }\textbf {\bibinfo {volume} {99}},\ \bibinfo
  {pages} {033625} (\bibinfo {year} {2019})}\BibitemShut {NoStop}%
\bibitem [{\citenamefont {Di~Carli}\ \emph {et~al.}(2019)\citenamefont
  {Di~Carli}, \citenamefont {Colquhoun}, \citenamefont {Henderson},
  \citenamefont {Flannigan}, \citenamefont {Oppo}, \citenamefont {Daley},
  \citenamefont {Kuhr},\ and\ \citenamefont {Haller}}]{DiCarliExcitation2019}%
  \BibitemOpen
  \bibfield  {author} {\bibinfo {author} {\bibfnamefont {Andrea}\ \bibnamefont
  {Di~Carli}}, \bibinfo {author} {\bibfnamefont {Craig~D.}\ \bibnamefont
  {Colquhoun}}, \bibinfo {author} {\bibfnamefont {Grant}\ \bibnamefont
  {Henderson}}, \bibinfo {author} {\bibfnamefont {Stuart}\ \bibnamefont
  {Flannigan}}, \bibinfo {author} {\bibfnamefont {Gian-Luca}\ \bibnamefont
  {Oppo}}, \bibinfo {author} {\bibfnamefont {Andrew~J.}\ \bibnamefont {Daley}},
  \bibinfo {author} {\bibfnamefont {Stefan}\ \bibnamefont {Kuhr}}, \ and\
  \bibinfo {author} {\bibfnamefont {Elmar}\ \bibnamefont {Haller}},\ }\bibfield
   {title} {\enquote {\bibinfo {title} {Excitation modes of bright matter-wave
  solitons},}\ }\href {\doibase 10.1103/PhysRevLett.123.123602} {\bibfield
  {journal} {\bibinfo  {journal} {Phys. Rev. Lett.}\ }\textbf {\bibinfo
  {volume} {123}},\ \bibinfo {pages} {123602} (\bibinfo {year}
  {2019})}\BibitemShut {NoStop}%
\bibitem [{sol()}]{solitary_wave}%
  \BibitemOpen
  \href@noop {} {}\bibinfo {note} {{Strictly, such realizations are in the form
  of bright solitary waves, as the integrability conditions necessary for true
  solitons are formally not fully satisfied.}}\BibitemShut {Stop}%
\bibitem [{\citenamefont {Veretenov}\ \emph {et~al.}(2007)\citenamefont
  {Veretenov}, \citenamefont {Rozhdestvensky}, \citenamefont {Rosanov},
  \citenamefont {Smirnov},\ and\ \citenamefont {Federov}}]{interf1}%
  \BibitemOpen
  \bibfield  {author} {\bibinfo {author} {\bibfnamefont {N.}~\bibnamefont
  {Veretenov}}, \bibinfo {author} {\bibfnamefont {Yu.}\ \bibnamefont
  {Rozhdestvensky}}, \bibinfo {author} {\bibfnamefont {N.}~\bibnamefont
  {Rosanov}}, \bibinfo {author} {\bibfnamefont {V.}~\bibnamefont {Smirnov}}, \
  and\ \bibinfo {author} {\bibfnamefont {S.}~\bibnamefont {Federov}},\
  }\bibfield  {title} {\enquote {\bibinfo {title} {Interferometric precision
  measurements with {B}ose--{E}instein condensate solitons formed by an optical
  lattice},}\ }\href {https://doi.org/10.1140/epjd/e2007-00129-2} {\bibfield
  {journal} {\bibinfo  {journal} {Eur. Phys. J. D.}\ }\textbf {\bibinfo
  {volume} {42}},\ \bibinfo {pages} {455} (\bibinfo {year} {2007})}\BibitemShut
  {NoStop}%
\bibitem [{\citenamefont {Abdullaev}\ and\ \citenamefont
  {Brazhnyi}(2012)}]{interf2}%
  \BibitemOpen
  \bibfield  {author} {\bibinfo {author} {\bibfnamefont {F.~Kh.}\ \bibnamefont
  {Abdullaev}}\ and\ \bibinfo {author} {\bibfnamefont {V.~A.}\ \bibnamefont
  {Brazhnyi}},\ }\bibfield  {title} {\enquote {\bibinfo {title} {Solitons in
  dipolar {B}ose-{E}instein condensates with a trap and barrier potential},}\
  }\href {\doibase 10.1088/0953-4075/45/8/085301} {\bibfield  {journal}
  {\bibinfo  {journal} {J. Phys. B: At. Mol. Opt. Phys.}\ }\textbf {\bibinfo
  {volume} {45}},\ \bibinfo {pages} {085301} (\bibinfo {year}
  {2012})}\BibitemShut {NoStop}%
\bibitem [{\citenamefont {Martin}\ and\ \citenamefont
  {Ruostekoski}(2012)}]{interf3}%
  \BibitemOpen
  \bibfield  {author} {\bibinfo {author} {\bibfnamefont {A.~D.}\ \bibnamefont
  {Martin}}\ and\ \bibinfo {author} {\bibfnamefont {J.}~\bibnamefont
  {Ruostekoski}},\ }\bibfield  {title} {\enquote {\bibinfo {title} {Quantum
  dynamics of atomic bright solitons under splitting and recollision, and
  implications for interferometry},}\ }\href {\doibase
  10.1088/1367-2630/14/4/043040} {\bibfield  {journal} {\bibinfo  {journal}
  {New J. Phys.}\ }\textbf {\bibinfo {volume} {14}},\ \bibinfo {pages} {043040}
  (\bibinfo {year} {2012})}\BibitemShut {NoStop}%
\bibitem [{\citenamefont {Helm}\ \emph {et~al.}(2012)\citenamefont {Helm},
  \citenamefont {Billam},\ and\ \citenamefont {Gardiner}}]{interf4}%
  \BibitemOpen
  \bibfield  {author} {\bibinfo {author} {\bibfnamefont {J.~L.}\ \bibnamefont
  {Helm}}, \bibinfo {author} {\bibfnamefont {T.~P.}\ \bibnamefont {Billam}}, \
  and\ \bibinfo {author} {\bibfnamefont {S.~A.}\ \bibnamefont {Gardiner}},\
  }\bibfield  {title} {\enquote {\bibinfo {title} {Bright matter-wave soliton
  collisions at narrow barriers},}\ }\href {\doibase
  10.1103/PhysRevA.85.053621} {\bibfield  {journal} {\bibinfo  {journal} {Phys.
  Rev. A}\ }\textbf {\bibinfo {volume} {85}},\ \bibinfo {pages} {053621}
  (\bibinfo {year} {2012})}\BibitemShut {NoStop}%
\bibitem [{\citenamefont {Polo}\ and\ \citenamefont
  {Ahufinger}(2013)}]{interf5}%
  \BibitemOpen
  \bibfield  {author} {\bibinfo {author} {\bibfnamefont {J.}~\bibnamefont
  {Polo}}\ and\ \bibinfo {author} {\bibfnamefont {V.}~\bibnamefont
  {Ahufinger}},\ }\bibfield  {title} {\enquote {\bibinfo {title} {Soliton-based
  matter-wave interferometer},}\ }\href {\doibase 10.1103/PhysRevA.88.053628}
  {\bibfield  {journal} {\bibinfo  {journal} {Phys. Rev. A}\ }\textbf {\bibinfo
  {volume} {88}},\ \bibinfo {pages} {053628} (\bibinfo {year}
  {2013})}\BibitemShut {NoStop}%
\bibitem [{\citenamefont {Cuevas}\ \emph {et~al.}(2013)\citenamefont {Cuevas},
  \citenamefont {Kevrekedis}, \citenamefont {Malomed}, \citenamefont {Dyke},\
  and\ \citenamefont {Hulet}}]{interf6}%
  \BibitemOpen
  \bibfield  {author} {\bibinfo {author} {\bibfnamefont {J.}~\bibnamefont
  {Cuevas}}, \bibinfo {author} {\bibfnamefont {P.~G.}\ \bibnamefont
  {Kevrekedis}}, \bibinfo {author} {\bibfnamefont {B.~A.}\ \bibnamefont
  {Malomed}}, \bibinfo {author} {\bibfnamefont {P.}~\bibnamefont {Dyke}}, \
  and\ \bibinfo {author} {\bibfnamefont {R.~G.}\ \bibnamefont {Hulet}},\
  }\bibfield  {title} {\enquote {\bibinfo {title} {Interactions of solitons
  with a {G}aussian barrier: splitting and recombination in
  quasi-one-dimensional and three-dimensional settings},}\ }\href {\doibase
  10.1088/1367-2630/15/6/063006} {\bibfield  {journal} {\bibinfo  {journal}
  {New J. Phys.}\ }\textbf {\bibinfo {volume} {15}},\ \bibinfo {pages} {063006}
  (\bibinfo {year} {2013})}\BibitemShut {NoStop}%
\bibitem [{\citenamefont {Helm}\ \emph {et~al.}(2015)\citenamefont {Helm},
  \citenamefont {Cornish},\ and\ \citenamefont {Gardiner}}]{interf7}%
  \BibitemOpen
  \bibfield  {author} {\bibinfo {author} {\bibfnamefont {J.~L.}\ \bibnamefont
  {Helm}}, \bibinfo {author} {\bibfnamefont {S.~L.}\ \bibnamefont {Cornish}}, \
  and\ \bibinfo {author} {\bibfnamefont {S.~A.}\ \bibnamefont {Gardiner}},\
  }\bibfield  {title} {\enquote {\bibinfo {title} {Sagnac interferometry using
  bright matter-wave solitons},}\ }\href {\doibase
  10.1103/PhysRevLett.114.134101} {\bibfield  {journal} {\bibinfo  {journal}
  {Phys. Rev. Lett.}\ }\textbf {\bibinfo {volume} {114}},\ \bibinfo {pages}
  {134101} (\bibinfo {year} {2015})}\BibitemShut {NoStop}%
\bibitem [{\citenamefont {Sakaguchi}\ and\ \citenamefont
  {Malomed}(2016)}]{interf8}%
  \BibitemOpen
  \bibfield  {author} {\bibinfo {author} {\bibfnamefont {H.}~\bibnamefont
  {Sakaguchi}}\ and\ \bibinfo {author} {\bibfnamefont {B.~A.}\ \bibnamefont
  {Malomed}},\ }\bibfield  {title} {\enquote {\bibinfo {title} {Matter-wave
  soliton interferometer based on a nonlinear splitter},}\ }\href {\doibase
  10.1088/1367-2630/18/2/025020} {\bibfield  {journal} {\bibinfo  {journal}
  {New J. Phys.}\ }\textbf {\bibinfo {volume} {18}},\ \bibinfo {pages} {025020}
  (\bibinfo {year} {2016})}\BibitemShut {NoStop}%
\bibitem [{\citenamefont {McDonald}\ \emph {et~al.}(2014)\citenamefont
  {McDonald}, \citenamefont {Kuhn}, \citenamefont {Hardman}, \citenamefont
  {Bennetts}, \citenamefont {Everitt}, \citenamefont {Altin}, \citenamefont
  {Debs}, \citenamefont {Close},\ and\ \citenamefont {Robins}}]{interf9}%
  \BibitemOpen
  \bibfield  {author} {\bibinfo {author} {\bibfnamefont {G.~D.}\ \bibnamefont
  {McDonald}}, \bibinfo {author} {\bibfnamefont {C.~C.~N.}\ \bibnamefont
  {Kuhn}}, \bibinfo {author} {\bibfnamefont {K.~S.}\ \bibnamefont {Hardman}},
  \bibinfo {author} {\bibfnamefont {S.}~\bibnamefont {Bennetts}}, \bibinfo
  {author} {\bibfnamefont {P.~J.}\ \bibnamefont {Everitt}}, \bibinfo {author}
  {\bibfnamefont {P.~A.}\ \bibnamefont {Altin}}, \bibinfo {author}
  {\bibfnamefont {J.~E.}\ \bibnamefont {Debs}}, \bibinfo {author}
  {\bibfnamefont {J.~D.}\ \bibnamefont {Close}}, \ and\ \bibinfo {author}
  {\bibfnamefont {N.~P.}\ \bibnamefont {Robins}},\ }\bibfield  {title}
  {\enquote {\bibinfo {title} {Bright solitonic matter-wave interferometer},}\
  }\href {\doibase 10.1103/PhysRevLett.113.013002} {\bibfield  {journal}
  {\bibinfo  {journal} {Phys. Rev. Lett.}\ }\textbf {\bibinfo {volume} {113}},\
  \bibinfo {pages} {013002} (\bibinfo {year} {2014})}\BibitemShut {NoStop}%
\bibitem [{\citenamefont {Haine}(2018)}]{HaineQuantum2018}%
  \BibitemOpen
  \bibfield  {author} {\bibinfo {author} {\bibfnamefont {S.~A.}\ \bibnamefont
  {Haine}},\ }\bibfield  {title} {\enquote {\bibinfo {title} {Quantum noise in
  bright soliton matterwave interferometry},}\ }\href {\doibase
  10.1088/1367-2630/aab47f} {\bibfield  {journal} {\bibinfo  {journal} {New J.
  Phys.}\ }\textbf {\bibinfo {volume} {20}},\ \bibinfo {pages} {033009}
  (\bibinfo {year} {2018})}\BibitemShut {NoStop}%
\bibitem [{\citenamefont {Nguyen}\ \emph {et~al.}(2014)\citenamefont {Nguyen},
  \citenamefont {Dyke}, \citenamefont {Luo}, \citenamefont {Malomed},\ and\
  \citenamefont {Hulet}}]{NguyenCollisions2014}%
  \BibitemOpen
  \bibfield  {author} {\bibinfo {author} {\bibfnamefont {J.~H.~V.}\
  \bibnamefont {Nguyen}}, \bibinfo {author} {\bibfnamefont {P.}~\bibnamefont
  {Dyke}}, \bibinfo {author} {\bibfnamefont {D.}~\bibnamefont {Luo}}, \bibinfo
  {author} {\bibfnamefont {B.~A.}\ \bibnamefont {Malomed}}, \ and\ \bibinfo
  {author} {\bibfnamefont {R.~G.}\ \bibnamefont {Hulet}},\ }\bibfield  {title}
  {\enquote {\bibinfo {title} {Collisions of matter-wave solitons},}\ }\href
  {\doibase 10.1038/nphys3135} {\bibfield  {journal} {\bibinfo  {journal} {Nat.
  Phys.}\ }\textbf {\bibinfo {volume} {10}},\ \bibinfo {pages} {918} (\bibinfo
  {year} {2014})}\BibitemShut {NoStop}%
\bibitem [{\citenamefont {{Wales}}\ \emph {et~al.}(2020)\citenamefont
  {{Wales}}, \citenamefont {{Rakonjac}}, \citenamefont {{Billam}},
  \citenamefont {{Helm}}, \citenamefont {{Gardiner}},\ and\ \citenamefont
  {{Cornish}}}]{exp_dur}%
  \BibitemOpen
  \bibfield  {author} {\bibinfo {author} {\bibfnamefont {O.~J.}\ \bibnamefont
  {{Wales}}}, \bibinfo {author} {\bibfnamefont {A.}~\bibnamefont {{Rakonjac}}},
  \bibinfo {author} {\bibfnamefont {T.~P.}\ \bibnamefont {{Billam}}}, \bibinfo
  {author} {\bibfnamefont {J.~L.}\ \bibnamefont {{Helm}}}, \bibinfo {author}
  {\bibfnamefont {S.~A.}\ \bibnamefont {{Gardiner}}}, \ and\ \bibinfo {author}
  {\bibfnamefont {S.~L.}\ \bibnamefont {{Cornish}}},\ }\bibfield  {title}
  {\enquote {\bibinfo {title} {{Splitting and recombination of
  bright-solitary-matter waves}},}\ }\href {\doibase
  https://doi.org/10.1038/s42005-020-0320-8} {\bibfield  {journal} {\bibinfo
  {journal} {Commun. Phys.}\ }\textbf {\bibinfo {volume} {3}},\ \bibinfo
  {pages} {51} (\bibinfo {year} {2020})}\BibitemShut {NoStop}%
\bibitem [{\citenamefont {Marchant}\ \emph {et~al.}(2013)\citenamefont
  {Marchant}, \citenamefont {Billam}, \citenamefont {Wiles}, \citenamefont
  {Yu}, \citenamefont {Gardiner},\ and\ \citenamefont
  {Cornish}}]{MarchantControlled2013}%
  \BibitemOpen
  \bibfield  {author} {\bibinfo {author} {\bibfnamefont {A.~L.}\ \bibnamefont
  {Marchant}}, \bibinfo {author} {\bibfnamefont {T.~P.}\ \bibnamefont
  {Billam}}, \bibinfo {author} {\bibfnamefont {T.~P.}\ \bibnamefont {Wiles}},
  \bibinfo {author} {\bibfnamefont {M.~M.~H.}\ \bibnamefont {Yu}}, \bibinfo
  {author} {\bibfnamefont {S.~A.}\ \bibnamefont {Gardiner}}, \ and\ \bibinfo
  {author} {\bibfnamefont {S.~L.}\ \bibnamefont {Cornish}},\ }\bibfield
  {title} {\enquote {\bibinfo {title} {Controlled formation and reflection of a
  bright solitary matter-wave},}\ }\href {\doibase 10.1038/ncomms2893}
  {\bibfield  {journal} {\bibinfo  {journal} {Nat. Commun.}\ }\textbf {\bibinfo
  {volume} {4}},\ \bibinfo {pages} {1865} (\bibinfo {year} {2013})}\BibitemShut
  {NoStop}%
\bibitem [{\citenamefont {\L{}\k{a}cki}\ \emph {et~al.}(2016)\citenamefont
  {\L{}\k{a}cki}, \citenamefont {Baranov}, \citenamefont {Pichler},\ and\
  \citenamefont {Zoller}}]{ds_tech_1}%
  \BibitemOpen
  \bibfield  {author} {\bibinfo {author} {\bibfnamefont {M.}~\bibnamefont
  {\L{}\k{a}cki}}, \bibinfo {author} {\bibfnamefont {M.~A.}\ \bibnamefont
  {Baranov}}, \bibinfo {author} {\bibfnamefont {H.}~\bibnamefont {Pichler}}, \
  and\ \bibinfo {author} {\bibfnamefont {P.}~\bibnamefont {Zoller}},\
  }\bibfield  {title} {\enquote {\bibinfo {title} {Nanoscale ``dark state''
  optical potentials for cold atoms},}\ }\href {\doibase
  10.1103/PhysRevLett.117.233001} {\bibfield  {journal} {\bibinfo  {journal}
  {Phys. Rev. Lett.}\ }\textbf {\bibinfo {volume} {117}},\ \bibinfo {pages}
  {233001} (\bibinfo {year} {2016})}\BibitemShut {NoStop}%
\bibitem [{\citenamefont {Jendrzejewski}\ \emph {et~al.}(2016)\citenamefont
  {Jendrzejewski}, \citenamefont {Eckel}, \citenamefont {Tiecke}, \citenamefont
  {Juzeli\ifmmode~\bar{u}\else \={u}\fi{}nas}, \citenamefont {Campbell},
  \citenamefont {Jiang},\ and\ \citenamefont {Gorshkov}}]{ds_tech_2}%
  \BibitemOpen
  \bibfield  {author} {\bibinfo {author} {\bibfnamefont {F.}~\bibnamefont
  {Jendrzejewski}}, \bibinfo {author} {\bibfnamefont {S.}~\bibnamefont
  {Eckel}}, \bibinfo {author} {\bibfnamefont {T.~G.}\ \bibnamefont {Tiecke}},
  \bibinfo {author} {\bibfnamefont {G.}~\bibnamefont
  {Juzeli\ifmmode~\bar{u}\else \={u}\fi{}nas}}, \bibinfo {author}
  {\bibfnamefont {G.~K.}\ \bibnamefont {Campbell}}, \bibinfo {author}
  {\bibfnamefont {Liang}\ \bibnamefont {Jiang}}, \ and\ \bibinfo {author}
  {\bibfnamefont {A.~V.}\ \bibnamefont {Gorshkov}},\ }\bibfield  {title}
  {\enquote {\bibinfo {title} {Subwavelength-width optical tunnel junctions for
  ultracold atoms},}\ }\href {\doibase 10.1103/PhysRevA.94.063422} {\bibfield
  {journal} {\bibinfo  {journal} {Phys. Rev. A}\ }\textbf {\bibinfo {volume}
  {94}},\ \bibinfo {pages} {063422} (\bibinfo {year} {2016})}\BibitemShut
  {NoStop}%
\bibitem [{\citenamefont {Ge}\ and\ \citenamefont {Zubairy}(2020)}]{ds_tech_3}%
  \BibitemOpen
  \bibfield  {author} {\bibinfo {author} {\bibfnamefont {W.}~\bibnamefont
  {Ge}}\ and\ \bibinfo {author} {\bibfnamefont {M.~S.}\ \bibnamefont
  {Zubairy}},\ }\bibfield  {title} {\enquote {\bibinfo {title} {Dark-state
  optical potential barriers with nanoscale spacing},}\ }\href {\doibase
  10.1103/PhysRevA.101.023403} {\bibfield  {journal} {\bibinfo  {journal}
  {Phys. Rev. A}\ }\textbf {\bibinfo {volume} {101}},\ \bibinfo {pages}
  {023403} (\bibinfo {year} {2020})}\BibitemShut {NoStop}%
\bibitem [{\citenamefont {Subhankar}\ \emph {et~al.}(2019)\citenamefont
  {Subhankar}, \citenamefont {Bienias}, \citenamefont {Titum}, \citenamefont
  {Tsui}, \citenamefont {Wang}, \citenamefont {Gorshkov}, \citenamefont
  {Rolston},\ and\ \citenamefont {Porto}}]{ds_tech_4}%
  \BibitemOpen
  \bibfield  {author} {\bibinfo {author} {\bibfnamefont {S.}~\bibnamefont
  {Subhankar}}, \bibinfo {author} {\bibfnamefont {P.}~\bibnamefont {Bienias}},
  \bibinfo {author} {\bibfnamefont {P.}~\bibnamefont {Titum}}, \bibinfo
  {author} {\bibfnamefont {T-C.}\ \bibnamefont {Tsui}}, \bibinfo {author}
  {\bibfnamefont {Y.}~\bibnamefont {Wang}}, \bibinfo {author} {\bibfnamefont
  {A.~V.}\ \bibnamefont {Gorshkov}}, \bibinfo {author} {\bibfnamefont {S.~L.}\
  \bibnamefont {Rolston}}, \ and\ \bibinfo {author} {\bibfnamefont {J.~V.}\
  \bibnamefont {Porto}},\ }\bibfield  {title} {\enquote {\bibinfo {title}
  {Floquet engineering of optical lattices with spatial features and
  periodicity below the diffraction limit},}\ }\href {\doibase
  10.1088/1367-2630/ab500f} {\bibfield  {journal} {\bibinfo  {journal} {New J.
  Phys.}\ }\textbf {\bibinfo {volume} {21}},\ \bibinfo {pages} {113058}
  (\bibinfo {year} {2019})}\BibitemShut {NoStop}%
\bibitem [{\citenamefont {Bienias}\ \emph {et~al.}(2020)\citenamefont
  {Bienias}, \citenamefont {Subhankar}, \citenamefont {Wang}, \citenamefont
  {Tsui}, \citenamefont {Jendrzejewski}, \citenamefont {Tiecke}, \citenamefont
  {Juzeli\ifmmode~\bar{u}\else \={u}\fi{}nas}, \citenamefont {Jiang},
  \citenamefont {Rolston}, \citenamefont {Porto},\ and\ \citenamefont
  {Gorshkov}}]{ds_tech_5}%
  \BibitemOpen
  \bibfield  {author} {\bibinfo {author} {\bibfnamefont {P.}~\bibnamefont
  {Bienias}}, \bibinfo {author} {\bibfnamefont {S.}~\bibnamefont {Subhankar}},
  \bibinfo {author} {\bibfnamefont {Y.}~\bibnamefont {Wang}}, \bibinfo {author}
  {\bibfnamefont {T-C.}\ \bibnamefont {Tsui}}, \bibinfo {author} {\bibfnamefont
  {F.}~\bibnamefont {Jendrzejewski}}, \bibinfo {author} {\bibfnamefont
  {T.}~\bibnamefont {Tiecke}}, \bibinfo {author} {\bibfnamefont
  {G.}~\bibnamefont {Juzeli\ifmmode~\bar{u}\else \={u}\fi{}nas}}, \bibinfo
  {author} {\bibfnamefont {L.}~\bibnamefont {Jiang}}, \bibinfo {author}
  {\bibfnamefont {S.~L.}\ \bibnamefont {Rolston}}, \bibinfo {author}
  {\bibfnamefont {J.~V.}\ \bibnamefont {Porto}}, \ and\ \bibinfo {author}
  {\bibfnamefont {A.~V.}\ \bibnamefont {Gorshkov}},\ }\bibfield  {title}
  {\enquote {\bibinfo {title} {Coherent optical nanotweezers for ultracold
  atoms},}\ }\href {\doibase 10.1103/PhysRevA.102.013306} {\bibfield  {journal}
  {\bibinfo  {journal} {Phys. Rev. A}\ }\textbf {\bibinfo {volume} {102}},\
  \bibinfo {pages} {013306} (\bibinfo {year} {2020})}\BibitemShut {NoStop}%
\bibitem [{\citenamefont {Tsui}\ \emph {et~al.}(2020)\citenamefont {Tsui},
  \citenamefont {Wang}, \citenamefont {Subhankar}, \citenamefont {Porto},\ and\
  \citenamefont {Rolston}}]{ds_tech_6}%
  \BibitemOpen
  \bibfield  {author} {\bibinfo {author} {\bibfnamefont {T-C.}\ \bibnamefont
  {Tsui}}, \bibinfo {author} {\bibfnamefont {Y.}~\bibnamefont {Wang}}, \bibinfo
  {author} {\bibfnamefont {S.}~\bibnamefont {Subhankar}}, \bibinfo {author}
  {\bibfnamefont {J.~V.}\ \bibnamefont {Porto}}, \ and\ \bibinfo {author}
  {\bibfnamefont {S.~L.}\ \bibnamefont {Rolston}},\ }\bibfield  {title}
  {\enquote {\bibinfo {title} {Realization of a stroboscopic optical lattice
  for cold atoms with subwavelength spacing},}\ }\href {\doibase
  10.1103/PhysRevA.101.041603} {\bibfield  {journal} {\bibinfo  {journal}
  {Phys. Rev. A}\ }\textbf {\bibinfo {volume} {101}},\ \bibinfo {pages}
  {041603} (\bibinfo {year} {2020})}\BibitemShut {NoStop}%
\bibitem [{\citenamefont {Kapale}\ and\ \citenamefont
  {Agarwal}(2010)}]{ds_tech_7}%
  \BibitemOpen
  \bibfield  {author} {\bibinfo {author} {\bibfnamefont {K.~T.}\ \bibnamefont
  {Kapale}}\ and\ \bibinfo {author} {\bibfnamefont {G.~S.}\ \bibnamefont
  {Agarwal}},\ }\bibfield  {title} {\enquote {\bibinfo {title} {Subnanoscale
  resolution for microscopy via coherent population trapping},}\ }\href
  {\doibase 10.1364/OL.35.002792} {\bibfield  {journal} {\bibinfo  {journal}
  {Opt. Lett.}\ }\textbf {\bibinfo {volume} {35}},\ \bibinfo {pages} {2792}
  (\bibinfo {year} {2010})}\BibitemShut {NoStop}%
\bibitem [{\citenamefont {Dum}\ and\ \citenamefont
  {Olshanii}(1996)}]{ds_tech_8}%
  \BibitemOpen
  \bibfield  {author} {\bibinfo {author} {\bibfnamefont {R.}~\bibnamefont
  {Dum}}\ and\ \bibinfo {author} {\bibfnamefont {M.}~\bibnamefont {Olshanii}},\
  }\bibfield  {title} {\enquote {\bibinfo {title} {Gauge structures in
  atom-laser interaction: Bloch oscillations in a dark lattice},}\ }\href
  {\doibase 10.1103/PhysRevLett.76.1788} {\bibfield  {journal} {\bibinfo
  {journal} {Phys. Rev. Lett.}\ }\textbf {\bibinfo {volume} {76}},\ \bibinfo
  {pages} {1788} (\bibinfo {year} {1996})}\BibitemShut {NoStop}%
\bibitem [{\citenamefont {Goldman}\ \emph {et~al.}(2014)\citenamefont
  {Goldman}, \citenamefont {Juzeli{\={u}}nas}, \citenamefont {{\"{O}}hberg},\
  and\ \citenamefont {Spielman}}]{ds_tech_9}%
  \BibitemOpen
  \bibfield  {author} {\bibinfo {author} {\bibfnamefont {N.}~\bibnamefont
  {Goldman}}, \bibinfo {author} {\bibfnamefont {G.}~\bibnamefont
  {Juzeli{\={u}}nas}}, \bibinfo {author} {\bibfnamefont {P.}~\bibnamefont
  {{\"{O}}hberg}}, \ and\ \bibinfo {author} {\bibfnamefont {I.~B.}\
  \bibnamefont {Spielman}},\ }\bibfield  {title} {\enquote {\bibinfo {title}
  {Light-induced gauge fields for ultracold atoms},}\ }\href {\doibase
  10.1088/0034-4885/77/12/126401} {\bibfield  {journal} {\bibinfo  {journal}
  {Rep. Prog. Phys.}\ }\textbf {\bibinfo {volume} {77}},\ \bibinfo {pages}
  {126401} (\bibinfo {year} {2014})}\BibitemShut {NoStop}%
\bibitem [{\citenamefont {Zhao}\ \emph {et~al.}(2020)\citenamefont {Zhao},
  \citenamefont {Wang}, \citenamefont {Zhuang},\ and\ \citenamefont
  {Li}}]{Zhao2020}%
  \BibitemOpen
  \bibfield  {author} {\bibinfo {author} {\bibfnamefont {Yang}\ \bibnamefont
  {Zhao}}, \bibinfo {author} {\bibfnamefont {Shaokai}\ \bibnamefont {Wang}},
  \bibinfo {author} {\bibfnamefont {Wei}\ \bibnamefont {Zhuang}}, \ and\
  \bibinfo {author} {\bibfnamefont {Tianchu}\ \bibnamefont {Li}},\ }\bibfield
  {title} {\enquote {\bibinfo {title} {Raman-laser system for absolute
  gravimeter based on $^{87}${Rb} atom interferometer},}\ }\href {\doibase
  10.3390/photonics7020032} {\bibfield  {journal} {\bibinfo  {journal}
  {Photonics}\ }\textbf {\bibinfo {volume} {7}},\ \bibinfo {pages} {32}
  (\bibinfo {year} {2020})}\BibitemShut {NoStop}%
\bibitem [{\citenamefont {Arias}\ \emph {et~al.}(2017)\citenamefont {Arias},
  \citenamefont {Abediyeh}, \citenamefont {Hamzeloui},\ and\ \citenamefont
  {Gomez}}]{Arias2017}%
  \BibitemOpen
  \bibfield  {author} {\bibinfo {author} {\bibfnamefont {N.}~\bibnamefont
  {Arias}}, \bibinfo {author} {\bibfnamefont {V.}~\bibnamefont {Abediyeh}},
  \bibinfo {author} {\bibfnamefont {S.}~\bibnamefont {Hamzeloui}}, \ and\
  \bibinfo {author} {\bibfnamefont {E.}~\bibnamefont {Gomez}},\ }\bibfield
  {title} {\enquote {\bibinfo {title} {Low phase noise beams for {R}aman
  transitions with a phase modulator and a highly birefringent crystal},}\
  }\href {\doibase 10.1364/OE.25.005290} {\bibfield  {journal} {\bibinfo
  {journal} {Opt. Express}\ }\textbf {\bibinfo {volume} {25}},\ \bibinfo
  {pages} {5290} (\bibinfo {year} {2017})}\BibitemShut {NoStop}%
\bibitem [{\citenamefont {Rosi}\ \emph {et~al.}(2014)\citenamefont {Rosi},
  \citenamefont {Sorrentino}, \citenamefont {Cacciapuoti}, \citenamefont
  {Prevedelli},\ and\ \citenamefont {Tino}}]{Rosi2014}%
  \BibitemOpen
  \bibfield  {author} {\bibinfo {author} {\bibfnamefont {G.}~\bibnamefont
  {Rosi}}, \bibinfo {author} {\bibfnamefont {F.}~\bibnamefont {Sorrentino}},
  \bibinfo {author} {\bibfnamefont {L.}~\bibnamefont {Cacciapuoti}}, \bibinfo
  {author} {\bibfnamefont {M.}~\bibnamefont {Prevedelli}}, \ and\ \bibinfo
  {author} {\bibfnamefont {G.~M.}\ \bibnamefont {Tino}},\ }\bibfield  {title}
  {\enquote {\bibinfo {title} {Precision measurement of the {N}ewtonian
  gravitational constant using cold atoms},}\ }\href {\doibase
  10.1038/nature13433} {\bibfield  {journal} {\bibinfo  {journal} {Nature}\
  }\textbf {\bibinfo {volume} {510}},\ \bibinfo {pages} {518} (\bibinfo {year}
  {2014})}\BibitemShut {NoStop}%
\bibitem [{\citenamefont {Meyrath}\ \emph {et~al.}(2005)\citenamefont
  {Meyrath}, \citenamefont {Schreck}, \citenamefont {Hanssen}, \citenamefont
  {Chuu},\ and\ \citenamefont {Raizen}}]{HG_trap}%
  \BibitemOpen
  \bibfield  {author} {\bibinfo {author} {\bibfnamefont {T.~P.}\ \bibnamefont
  {Meyrath}}, \bibinfo {author} {\bibfnamefont {F.}~\bibnamefont {Schreck}},
  \bibinfo {author} {\bibfnamefont {J.~L.}\ \bibnamefont {Hanssen}}, \bibinfo
  {author} {\bibfnamefont {C.~S.}\ \bibnamefont {Chuu}}, \ and\ \bibinfo
  {author} {\bibfnamefont {M.~G.}\ \bibnamefont {Raizen}},\ }\bibfield  {title}
  {\enquote {\bibinfo {title} {A high frequency optical trap for atoms using
  {H}ermite--{G}aussian beams},}\ }\href {\doibase 10.1364/OPEX.13.002843}
  {\bibfield  {journal} {\bibinfo  {journal} {Opt. Express}\ }\textbf {\bibinfo
  {volume} {13}},\ \bibinfo {pages} {2843} (\bibinfo {year}
  {2005})}\BibitemShut {NoStop}%
\bibitem [{\citenamefont {Zupancic}\ \emph {et~al.}(2016)\citenamefont
  {Zupancic}, \citenamefont {Preiss}, \citenamefont {Ma}, \citenamefont
  {Lukin}, \citenamefont {Tai}, \citenamefont {Rispoli}, \citenamefont
  {Islam},\ and\ \citenamefont {Greiner}}]{Zupancic2016}%
  \BibitemOpen
  \bibfield  {author} {\bibinfo {author} {\bibfnamefont {Philip}\ \bibnamefont
  {Zupancic}}, \bibinfo {author} {\bibfnamefont {Philipp~M.}\ \bibnamefont
  {Preiss}}, \bibinfo {author} {\bibfnamefont {Ruichao}\ \bibnamefont {Ma}},
  \bibinfo {author} {\bibfnamefont {Alexander}\ \bibnamefont {Lukin}}, \bibinfo
  {author} {\bibfnamefont {M.~Eric}\ \bibnamefont {Tai}}, \bibinfo {author}
  {\bibfnamefont {Matthew}\ \bibnamefont {Rispoli}}, \bibinfo {author}
  {\bibfnamefont {Rajibul}\ \bibnamefont {Islam}}, \ and\ \bibinfo {author}
  {\bibfnamefont {Markus}\ \bibnamefont {Greiner}},\ }\bibfield  {title}
  {\enquote {\bibinfo {title} {Ultra-precise holographic beam shaping for
  microscopic quantum control},}\ }\href {\doibase 10.1364/OE.24.013881}
  {\bibfield  {journal} {\bibinfo  {journal} {Opt. Express}\ }\textbf {\bibinfo
  {volume} {24}},\ \bibinfo {pages} {13881} (\bibinfo {year}
  {2016})}\BibitemShut {NoStop}%
\bibitem [{\citenamefont {Uehlinger}\ \emph {et~al.}(2013)\citenamefont
  {Uehlinger}, \citenamefont {Jotzu}, \citenamefont {Messer}, \citenamefont
  {Greif}, \citenamefont {Hofstetter}, \citenamefont {Bissbort},\ and\
  \citenamefont {Esslinger}}]{Uehlinger2013}%
  \BibitemOpen
  \bibfield  {author} {\bibinfo {author} {\bibfnamefont {Thomas}\ \bibnamefont
  {Uehlinger}}, \bibinfo {author} {\bibfnamefont {Gregor}\ \bibnamefont
  {Jotzu}}, \bibinfo {author} {\bibfnamefont {Michael}\ \bibnamefont {Messer}},
  \bibinfo {author} {\bibfnamefont {Daniel}\ \bibnamefont {Greif}}, \bibinfo
  {author} {\bibfnamefont {Walter}\ \bibnamefont {Hofstetter}}, \bibinfo
  {author} {\bibfnamefont {Ulf}\ \bibnamefont {Bissbort}}, \ and\ \bibinfo
  {author} {\bibfnamefont {Tilman}\ \bibnamefont {Esslinger}},\ }\bibfield
  {title} {\enquote {\bibinfo {title} {Artificial graphene with tunable
  interactions},}\ }\href {\doibase 10.1103/PhysRevLett.111.185307} {\bibfield
  {journal} {\bibinfo  {journal} {Phys. Rev. Lett.}\ }\textbf {\bibinfo
  {volume} {111}},\ \bibinfo {pages} {185307} (\bibinfo {year}
  {2013})}\BibitemShut {NoStop}%
\bibitem [{\citenamefont {Gati}\ \emph {et~al.}(2006)\citenamefont {Gati},
  \citenamefont {Albiez}, \citenamefont {F\"{o}lling}, \citenamefont
  {Hemmerling},\ and\ \citenamefont {Oberthaler}}]{as_trap}%
  \BibitemOpen
  \bibfield  {author} {\bibinfo {author} {\bibfnamefont {R.}~\bibnamefont
  {Gati}}, \bibinfo {author} {\bibfnamefont {M.}~\bibnamefont {Albiez}},
  \bibinfo {author} {\bibfnamefont {J.}~\bibnamefont {F\"{o}lling}}, \bibinfo
  {author} {\bibfnamefont {B.}~\bibnamefont {Hemmerling}}, \ and\ \bibinfo
  {author} {\bibfnamefont {M.~K.}\ \bibnamefont {Oberthaler}},\ }\bibfield
  {title} {\enquote {\bibinfo {title} {Realization of a single {J}osephson
  junction for {B}ose--{E}instein condensates},}\ }\href {\doibase
  10.1007/s00340-005-2059-z10.1007/s00340-005-2059-z} {\bibfield  {journal}
  {\bibinfo  {journal} {App. Phys. B}\ }\textbf {\bibinfo {volume} {82}},\
  \bibinfo {pages} {207} (\bibinfo {year} {2006})}\BibitemShut {NoStop}%
\bibitem [{\citenamefont {Malomed}\ and\ \citenamefont
  {Tasgal}(1998)}]{int_vib}%
  \BibitemOpen
  \bibfield  {author} {\bibinfo {author} {\bibfnamefont {B.~A.}\ \bibnamefont
  {Malomed}}\ and\ \bibinfo {author} {\bibfnamefont {R.~S.}\ \bibnamefont
  {Tasgal}},\ }\bibfield  {title} {\enquote {\bibinfo {title} {Internal
  vibrations of a vector soliton in the coupled nonlinear {S}chr\"odinger
  equations},}\ }\href {\doibase 10.1103/PhysRevE.58.2564} {\bibfield
  {journal} {\bibinfo  {journal} {Phys. Rev. E}\ }\textbf {\bibinfo {volume}
  {58}},\ \bibinfo {pages} {2564} (\bibinfo {year} {1998})}\BibitemShut
  {NoStop}%
\bibitem [{\citenamefont {Manakov}(1974)}]{manakov1}%
  \BibitemOpen
  \bibfield  {author} {\bibinfo {author} {\bibfnamefont {S.~V.}\ \bibnamefont
  {Manakov}},\ }\bibfield  {title} {\enquote {\bibinfo {title} {On the theory
  of two-dimensional stationary self-focussing of electromagnetic waves},}\
  }\href {http://www.jetp.ac.ru/cgi-bin/e/index/e/38/2/p248?a=list} {\bibfield
  {journal} {\bibinfo  {journal} {Sov. Phys. JETP}\ }\textbf {\bibinfo {volume}
  {38}},\ \bibinfo {pages} {248} (\bibinfo {year} {1974})},\ \bibinfo {note}
  {[Zh. Eksp. Teor. Fiz. \textbf{65} 505, (1974)]}\BibitemShut {NoStop}%
\bibitem [{\citenamefont {Roberts}\ \emph {et~al.}(1998)\citenamefont
  {Roberts}, \citenamefont {Claussen}, \citenamefont {Burke}, \citenamefont
  {Greene}, \citenamefont {Cornell},\ and\ \citenamefont
  {Wieman}}]{feshbachresonance_1}%
  \BibitemOpen
  \bibfield  {author} {\bibinfo {author} {\bibfnamefont {J.~L.}\ \bibnamefont
  {Roberts}}, \bibinfo {author} {\bibfnamefont {N.~R.}\ \bibnamefont
  {Claussen}}, \bibinfo {author} {\bibfnamefont {J.~P.}\ \bibnamefont {Burke}},
  \bibinfo {author} {\bibfnamefont {C.~H.}\ \bibnamefont {Greene}}, \bibinfo
  {author} {\bibfnamefont {E.~A.}\ \bibnamefont {Cornell}}, \ and\ \bibinfo
  {author} {\bibfnamefont {C.~E.}\ \bibnamefont {Wieman}},\ }\bibfield  {title}
  {\enquote {\bibinfo {title} {Resonant magnetic field control of elastic
  scattering in cold $^{85}${R}b},}\ }\href {\doibase
  10.1103/PhysRevLett.81.5109} {\bibfield  {journal} {\bibinfo  {journal}
  {Phys. Rev. Lett.}\ }\textbf {\bibinfo {volume} {81}},\ \bibinfo {pages}
  {5109} (\bibinfo {year} {1998})}\BibitemShut {NoStop}%
\bibitem [{\citenamefont {Blackley}\ \emph {et~al.}(2013)\citenamefont
  {Blackley}, \citenamefont {Le~Sueur}, \citenamefont {Hutson}, \citenamefont
  {McCarron}, \citenamefont {K\"oppinger}, \citenamefont {Cho}, \citenamefont
  {Jenkin},\ and\ \citenamefont {Cornish}}]{feshbachresonance_2}%
  \BibitemOpen
  \bibfield  {author} {\bibinfo {author} {\bibfnamefont {C.~L.}\ \bibnamefont
  {Blackley}}, \bibinfo {author} {\bibfnamefont {C.~R}\ \bibnamefont
  {Le~Sueur}}, \bibinfo {author} {\bibfnamefont {J.~M.}\ \bibnamefont
  {Hutson}}, \bibinfo {author} {\bibfnamefont {D.~J.}\ \bibnamefont
  {McCarron}}, \bibinfo {author} {\bibfnamefont {M.~P.}\ \bibnamefont
  {K\"oppinger}}, \bibinfo {author} {\bibfnamefont {H.~W.}\ \bibnamefont
  {Cho}}, \bibinfo {author} {\bibfnamefont {D.~L.}\ \bibnamefont {Jenkin}}, \
  and\ \bibinfo {author} {\bibfnamefont {S.~L.}\ \bibnamefont {Cornish}},\
  }\bibfield  {title} {\enquote {\bibinfo {title} {Feshbach resonances in
  ultracold ${}^{85}${R}b},}\ }\href {\doibase 10.1103/PhysRevA.87.033611}
  {\bibfield  {journal} {\bibinfo  {journal} {Phys. Rev. A}\ }\textbf {\bibinfo
  {volume} {87}},\ \bibinfo {pages} {033611} (\bibinfo {year}
  {2013})}\BibitemShut {NoStop}%
\bibitem [{\citenamefont {Ruprecht}\ \emph {et~al.}(1995)\citenamefont
  {Ruprecht}, \citenamefont {Holland}, \citenamefont {Burnett},\ and\
  \citenamefont {Edwards}}]{stab_2}%
  \BibitemOpen
  \bibfield  {author} {\bibinfo {author} {\bibfnamefont {P.~A.}\ \bibnamefont
  {Ruprecht}}, \bibinfo {author} {\bibfnamefont {M.~J.}\ \bibnamefont
  {Holland}}, \bibinfo {author} {\bibfnamefont {K.}~\bibnamefont {Burnett}}, \
  and\ \bibinfo {author} {\bibfnamefont {M.}~\bibnamefont {Edwards}},\
  }\bibfield  {title} {\enquote {\bibinfo {title} {Time-dependent solution of
  the nonlinear {S}chr\"odinger equation for {B}ose-condensed trapped neutral
  atoms},}\ }\href {\doibase 10.1103/PhysRevA.51.4704} {\bibfield  {journal}
  {\bibinfo  {journal} {Phys. Rev. A}\ }\textbf {\bibinfo {volume} {51}},\
  \bibinfo {pages} {4704} (\bibinfo {year} {1995})}\BibitemShut {NoStop}%
\bibitem [{\citenamefont {Roberts}\ \emph {et~al.}(2001)\citenamefont
  {Roberts}, \citenamefont {Claussen}, \citenamefont {Cornish}, \citenamefont
  {Donley}, \citenamefont {Cornell},\ and\ \citenamefont {Wieman}}]{stab_3}%
  \BibitemOpen
  \bibfield  {author} {\bibinfo {author} {\bibfnamefont {J.~L.}\ \bibnamefont
  {Roberts}}, \bibinfo {author} {\bibfnamefont {N.~R.}\ \bibnamefont
  {Claussen}}, \bibinfo {author} {\bibfnamefont {S.~L.}\ \bibnamefont
  {Cornish}}, \bibinfo {author} {\bibfnamefont {E.~A.}\ \bibnamefont {Donley}},
  \bibinfo {author} {\bibfnamefont {E.~A.}\ \bibnamefont {Cornell}}, \ and\
  \bibinfo {author} {\bibfnamefont {C.~E.}\ \bibnamefont {Wieman}},\ }\bibfield
   {title} {\enquote {\bibinfo {title} {Controlled collapse of a
  {B}ose--{E}instein condensate},}\ }\href {\doibase
  10.1103/PhysRevLett.86.4211} {\bibfield  {journal} {\bibinfo  {journal}
  {Phys. Rev. Lett.}\ }\textbf {\bibinfo {volume} {86}},\ \bibinfo {pages}
  {4211} (\bibinfo {year} {2001})}\BibitemShut {NoStop}%
\bibitem [{dim()}]{dimensionless_variables}%
  \BibitemOpen
  \href@noop {} {}\bibinfo {note} {{Heuristically this can be expressed as
  $\hbar = m= |g^{1\mathrm{D}}_{11}|N= 1$.}}\BibitemShut {Stop}%
\bibitem [{\citenamefont {Landau}\ and\ \citenamefont
  {Lifshitz}(1959)}]{landauandlifshitz}%
  \BibitemOpen
  \bibfield  {author} {\bibinfo {author} {\bibfnamefont {L.~D.}\ \bibnamefont
  {Landau}}\ and\ \bibinfo {author} {\bibfnamefont {E.~M.}\ \bibnamefont
  {Lifshitz}},\ }\href@noop {} {\emph {\bibinfo {title} {Quantum Mechanics
  (Non-Relativistic Theory)}}}\ (\bibinfo  {publisher} {Pergamon},\ \bibinfo
  {year} {1959})\BibitemShut {NoStop}%
\bibitem [{\citenamefont {Holmer}\ \emph {et~al.}(2007)\citenamefont {Holmer},
  \citenamefont {Marzuola},\ and\ \citenamefont {Zworski}}]{dfanalytics1}%
  \BibitemOpen
  \bibfield  {author} {\bibinfo {author} {\bibfnamefont {J.}~\bibnamefont
  {Holmer}}, \bibinfo {author} {\bibfnamefont {J.}~\bibnamefont {Marzuola}}, \
  and\ \bibinfo {author} {\bibfnamefont {M.}~\bibnamefont {Zworski}},\
  }\bibfield  {title} {\enquote {\bibinfo {title} {Fast soliton scattering by
  delta impurities},}\ }\href {https://doi.org/10.1007/s00220-007-0261-z}
  {\bibfield  {journal} {\bibinfo  {journal} {Commun. Math. Phys.}\ }\textbf
  {\bibinfo {volume} {274}},\ \bibinfo {pages} {187} (\bibinfo {year}
  {2007})}\BibitemShut {NoStop}%
\bibitem [{\citenamefont {Manju}\ \emph {et~al.}(2018)\citenamefont {Manju},
  \citenamefont {Hardman}, \citenamefont {Sooriyabandara}, \citenamefont
  {Wigley}, \citenamefont {Close}, \citenamefont {Robins}, \citenamefont
  {Hush},\ and\ \citenamefont {Szigeti}}]{tunneling2}%
  \BibitemOpen
  \bibfield  {author} {\bibinfo {author} {\bibfnamefont {P.}~\bibnamefont
  {Manju}}, \bibinfo {author} {\bibfnamefont {K.~S.}\ \bibnamefont {Hardman}},
  \bibinfo {author} {\bibfnamefont {M.~A.}\ \bibnamefont {Sooriyabandara}},
  \bibinfo {author} {\bibfnamefont {P.~B.}\ \bibnamefont {Wigley}}, \bibinfo
  {author} {\bibfnamefont {J.~D.}\ \bibnamefont {Close}}, \bibinfo {author}
  {\bibfnamefont {N.~P.}\ \bibnamefont {Robins}}, \bibinfo {author}
  {\bibfnamefont {M.~R.}\ \bibnamefont {Hush}}, \ and\ \bibinfo {author}
  {\bibfnamefont {S.~S.}\ \bibnamefont {Szigeti}},\ }\bibfield  {title}
  {\enquote {\bibinfo {title} {Quantum tunneling dynamics of an interacting
  {B}ose--{E}instein condensate through a {G}aussian barrier},}\ }\href
  {\doibase https://doi.org/10.1103/PhysRevA.98.053629} {\bibfield  {journal}
  {\bibinfo  {journal} {Phys. Rev. A}\ }\textbf {\bibinfo {volume} {98}},\
  \bibinfo {pages} {053629} (\bibinfo {year} {2018})}\BibitemShut {NoStop}%
\bibitem [{\citenamefont {Helm}\ \emph {et~al.}(2014)\citenamefont {Helm},
  \citenamefont {Rooney}, \citenamefont {Weiss},\ and\ \citenamefont
  {Gardiner}}]{si1}%
  \BibitemOpen
  \bibfield  {author} {\bibinfo {author} {\bibfnamefont {J.~L.}\ \bibnamefont
  {Helm}}, \bibinfo {author} {\bibfnamefont {S.~J.}\ \bibnamefont {Rooney}},
  \bibinfo {author} {\bibfnamefont {C.}~\bibnamefont {Weiss}}, \ and\ \bibinfo
  {author} {\bibfnamefont {S.~A.}\ \bibnamefont {Gardiner}},\ }\bibfield
  {title} {\enquote {\bibinfo {title} {Splitting bright matter-wave solitons on
  narrow potential barriers: quantum to classical transition and applications
  to interferometry},}\ }\href {\doibase 10.1103/PhysRevA.89.033610} {\bibfield
   {journal} {\bibinfo  {journal} {Phys. Rev. A}\ }\textbf {\bibinfo {volume}
  {89}},\ \bibinfo {pages} {033610} (\bibinfo {year} {2014})}\BibitemShut
  {NoStop}%
\bibitem [{fit()}]{fitting}%
  \BibitemOpen
  \href@noop {} {}\bibinfo {note} {{In detail, using
  \texttt{scipy.optimize.curve\_fit} \cite{SciPy2020}, we perform a
  least-squares fit to the numerical $T_2$ data assuming equal uncertainty in
  each data point, and plot one standard deviation uncertainties in the fit
  parameters obtained from the covariance matrix after scaling the minimized
  reduced $\chi^2$ statistic to $1$.}}\BibitemShut {Stop}%
\bibitem [{\citenamefont {Virtanen}\ \emph {et~al.}(2020)\citenamefont
  {Virtanen}, \citenamefont {Gommers}, \citenamefont {Oliphant}, \citenamefont
  {Haberland}, \citenamefont {Reddy}, \citenamefont {Cournapeau}, \citenamefont
  {Burovski}, \citenamefont {Peterson}, \citenamefont {Weckesser},
  \citenamefont {Bright}, \citenamefont {{van der Walt}}, \citenamefont
  {Brett}, \citenamefont {Wilson}, \citenamefont {Millman}, \citenamefont
  {Mayorov}, \citenamefont {Nelson}, \citenamefont {Jones}, \citenamefont
  {Kern}, \citenamefont {Larson}, \citenamefont {Carey}, \citenamefont {Polat},
  \citenamefont {Feng}, \citenamefont {Moore}, \citenamefont {{VanderPlas}},
  \citenamefont {Laxalde}, \citenamefont {Perktold}, \citenamefont {Cimrman},
  \citenamefont {Henriksen}, \citenamefont {Quintero}, \citenamefont {Harris},
  \citenamefont {Archibald}, \citenamefont {Ribeiro}, \citenamefont
  {Pedregosa}, \citenamefont {{van Mulbregt}},\ and\ \citenamefont {{SciPy 1.0
  Contributors}}}]{SciPy2020}%
  \BibitemOpen
  \bibfield  {author} {\bibinfo {author} {\bibfnamefont {P.}~\bibnamefont
  {Virtanen}}, \bibinfo {author} {\bibfnamefont {R.}~\bibnamefont {Gommers}},
  \bibinfo {author} {\bibfnamefont {T.~E.}\ \bibnamefont {Oliphant}}, \bibinfo
  {author} {\bibfnamefont {M.}~\bibnamefont {Haberland}}, \bibinfo {author}
  {\bibfnamefont {T.}~\bibnamefont {Reddy}}, \bibinfo {author} {\bibfnamefont
  {D.}~\bibnamefont {Cournapeau}}, \bibinfo {author} {\bibfnamefont
  {E.}~\bibnamefont {Burovski}}, \bibinfo {author} {\bibfnamefont
  {P.}~\bibnamefont {Peterson}}, \bibinfo {author} {\bibfnamefont
  {W.}~\bibnamefont {Weckesser}}, \bibinfo {author} {\bibfnamefont
  {J.}~\bibnamefont {Bright}}, \bibinfo {author} {\bibfnamefont {S.~J.}\
  \bibnamefont {{van der Walt}}}, \bibinfo {author} {\bibfnamefont
  {M.}~\bibnamefont {Brett}}, \bibinfo {author} {\bibfnamefont
  {J.}~\bibnamefont {Wilson}}, \bibinfo {author} {\bibfnamefont {K.~J.}\
  \bibnamefont {Millman}}, \bibinfo {author} {\bibfnamefont {N.}~\bibnamefont
  {Mayorov}}, \bibinfo {author} {\bibfnamefont {A.~R.~J.}\ \bibnamefont
  {Nelson}}, \bibinfo {author} {\bibfnamefont {E.}~\bibnamefont {Jones}},
  \bibinfo {author} {\bibfnamefont {R.}~\bibnamefont {Kern}}, \bibinfo {author}
  {\bibfnamefont {E.}~\bibnamefont {Larson}}, \bibinfo {author} {\bibfnamefont
  {C.~J.}\ \bibnamefont {Carey}}, \bibinfo {author} {\bibfnamefont
  {{\.I}.}~\bibnamefont {Polat}}, \bibinfo {author} {\bibfnamefont
  {Y.}~\bibnamefont {Feng}}, \bibinfo {author} {\bibfnamefont {E.~W.}\
  \bibnamefont {Moore}}, \bibinfo {author} {\bibfnamefont {J.}~\bibnamefont
  {{VanderPlas}}}, \bibinfo {author} {\bibfnamefont {D.}~\bibnamefont
  {Laxalde}}, \bibinfo {author} {\bibfnamefont {J.}~\bibnamefont {Perktold}},
  \bibinfo {author} {\bibfnamefont {R.}~\bibnamefont {Cimrman}}, \bibinfo
  {author} {\bibfnamefont {I.}~\bibnamefont {Henriksen}}, \bibinfo {author}
  {\bibfnamefont {E.~A.}\ \bibnamefont {Quintero}}, \bibinfo {author}
  {\bibfnamefont {C.~R.}\ \bibnamefont {Harris}}, \bibinfo {author}
  {\bibfnamefont {A.~M.}\ \bibnamefont {Archibald}}, \bibinfo {author}
  {\bibfnamefont {A.~H.}\ \bibnamefont {Ribeiro}}, \bibinfo {author}
  {\bibfnamefont {F.}~\bibnamefont {Pedregosa}}, \bibinfo {author}
  {\bibfnamefont {P.}~\bibnamefont {{van Mulbregt}}}, \ and\ \bibinfo {author}
  {\bibnamefont {{SciPy 1.0 Contributors}}},\ }\bibfield  {title} {\enquote
  {\bibinfo {title} {{{SciPy} 1.0: Fundamental Algorithms for Scientific
  Computing in Python}},}\ }\href {\doibase 10.1038/s41592-019-0686-2}
  {\bibfield  {journal} {\bibinfo  {journal} {Nature Methods}\ }\textbf
  {\bibinfo {volume} {17}},\ \bibinfo {pages} {261} (\bibinfo {year}
  {2020})}\BibitemShut {NoStop}%
\bibitem [{Note1()}]{Note1}%
  \BibitemOpen
  \bibinfo {note} {Specifically, the wavenumber $k$ required for equal
  splitting shifts by $\approx 2\protect \tmspace +\thinmuskip {.1667em}$\%,
  and $\approx 2\protect \tmspace +\thinmuskip {.1667em}$\% of the incoming
  atom flux is lost at this wavenumber.}\BibitemShut {Stop}%
\bibitem [{dat()}]{data}%
  \BibitemOpen
  \href {\doibase 10.15128/r11j92g749r} {\enquote {\bibinfo {title} {Data are
  available through {D}urham {U}niversity data management},}\ }\bibinfo
  {howpublished} {DOI:10.15128/r11j92g749r}\BibitemShut {NoStop}%
\end{thebibliography}%

\end{document}